\documentclass[a4paper,12pt]{article}
\pdfoutput=1
\usepackage{cite}
\usepackage{amsmath}
\usepackage{amsfonts}
\usepackage{amssymb}
\usepackage{graphicx, rotating}
\usepackage{epsfig}
\usepackage{latexsym}
\usepackage{graphicx}
\usepackage{color}
\usepackage{amsmath,bm,amssymb}
\usepackage{color}
\usepackage[english]{babel}
\usepackage{hyperref}

\renewcommand{\figurename}{Fig.}
\renewcommand{\tablename}{Table.}
\newcommand{\Slash}[1]{{\ooalign{\hfil#1\hfil\crcr\raise.167ex\hbox{/}}}}

\newcommand{\beq}{\begin{equation}}  \newcommand{\eeq}{\end{equation}}
\newcommand{\bef}{\begin{figure}}  \newcommand{\eef}{\end{figure}}
\newcommand{\bec}{\begin{center}}  \newcommand{\eec}{\end{center}}
  
\newcommand{\laq}[1]{\label{eq:#1}}  

\newcommand{\Eq}[1]{Eq.~(\ref{eq:#1})}
\newcommand{\Eqs}[1]{Eqs.~(\ref{eq:#1})}
\newcommand{\eq}[1]{(\ref{eq:#1})}
\newcommand{\Sec}[1]{Sec.~\ref{chap:#1}}
\newcommand{\ab}[1]{\left|{#1}\right|}
\newcommand{\vev}[1]{ \left\langle {#1} \right\rangle }

\newcommand{\lac}[1]{\label{chap:#1}}
\newcommand{\SU}[1]{{\rm SU{#1} } }

\def\({\left(}
\def\){\right)}

\def\O{\mathcal{O}}
\def\U{\mathop{\rm U}}

\newcommand{\OR}{~{\rm or}~}
\newcommand{\AND}{~{\rm and}~}
\newcommand{\EV}{ {\rm \, eV} }
\newcommand{\KEV}{ {\rm \, keV} }
\newcommand{\MEV}{ {\rm \, MeV} }
\newcommand{\GEV}{ {\rm \, GeV} }
\newcommand{\TEV}{ {\rm \, TeV} }

\def\a{\alpha}

\def\d{\delta}

\def\f{\phi}
\def\g{\gamma}
\def\h{\theta}

\def\l{\lambda}
\def\m{\mu}

\def\p{\psi}

\def\s{\sigma}
\def\t{\tau}

\def\x{\xi}

\def\z{\zeta}
\def\D{\Delta}
\def\G{\Gamma}
\def\H{\Theta}
\def\L{\Lambda}
\def\F{\Phi}

\def\tl{\tilde}
\def\*{\dagger}

\begin{document}
\begin{flushright}
\hspace{3cm} 
TU-1179
\end{flushright}
\vspace{.6cm}

\begin{center}

\vspace{1.5cm}
{\Large\bf  Hadrophobic Axion from GUT }
\vspace{1.5cm}

{\bf Fuminobu Takahashi$^{1,2}$} and {\bf Wen Yin$^{1}$}

\vspace{12pt}
\vspace{1.5cm}
{\em 

$^{1}$Department of Physics, Tohoku University,  
Sendai, Miyagi 980-8578, Japan \\
$^{2}$Kavli Institute for the Physics and Mathematics of the Universe (WPI),
University of Tokyo, Kashiwa 277--8583, Japan  \vspace{5pt}}

\vspace{1.5cm}
\abstract{
We propose a new kind of axion model derived from the Grand Unified Theory (GUT) based on $\SU(5)\times \U(1)_{\rm PQ}$. We demonstrate that, given a certain charge assignment and potential flavor models, the axion is naturally hadrophobic and provides a novel explanation for the required condition using isospin symmetry. {This axion can be the QCD axion that solves the strong CP problem.} 
Furthermore, to satisfy the limit on the axion-electron coupling from the tip of the red giant branch, we impose the condition of electrophobia to determine a possible PQ charge assignment consistent with GUT. We then discuss the possibility that the hadrophobic and electrophobic axion serves as an inflaton and dark matter, as in the ALP miracle scenario.
Interestingly, in the viable parameter region, the strong CP phase must be suppressed, providing another solution to the strong CP problem. This scenario is intimately linked to flavor physics, dark matter searches, and stellar cooling. Detecting such an axion with peculiar couplings in various experiments would serve as a probe for GUT and the origin of flavor.
}

\end{center}
\clearpage

\setcounter{page}{1}
\setcounter{footnote}{0}

\section{Introduction}

The QCD axion and axion-like particles (ALPs), collectively referred to as axions, are intriguing candidates for light dark matter (DM)~\cite{Preskill:1982cy,Abbott:1982af,Dine:1982ah} 
(see
Refs.\,\cite{Jaeckel:2010ni,Ringwald:2012hr,Arias:2012az,Graham:2015ouw,Marsh:2015xka,Irastorza:2018dyq,DiLuzio:2020wdo} for reviews).  
The axion is a pseudo Nambu-Goldstone boson associated with the spontaneously broken Peccei-Quinn (PQ) symmetry ~\cite{Peccei:1977hh,Peccei:1977ur,Weinberg:1977ma,Wilczek:1977pj}, which is anomalous under the standard model (SM) gauge groups. 
Axion models are frequently explored in the context of grand unified theory (GUT), which accounts for the charge quantization of SM particle contents (see, for example, recent studies of the QCD axion and GUT in Refs.~\cite{Ernst:2018bib, Lee:2018yak, FileviezPerez:2019ssf, Contino:2021ayn, Antusch:2023jok}). In fact, known quarks and leptons are neatly embedded in complete GUT multiplets.
In this paper we focus on the PQ extension of the $\SU(5)$ GUT 
with a symmetry of \beq \laq{sym}\SU(5)\times \U(1)_{\rm PQ}.\eeq  
Here $\SU(5)$ represents the GUT gauge symmetry and $\U(1)_{\rm PQ}$ denotes the global PQ symmetry,
with both assumed to commute with each other.

The QCD axion offers a promising solution to the strong CP problem~\cite{Peccei:1977hh,Peccei:1977ur,Weinberg:1977ma,Wilczek:1977pj}. The strength 
of its interactions is characterized by the decay constant $f_a$, conventionally normalized by its coupling to gluons. 
The  classical axion window on $f_a$ is given by $10^{8-9}\GEV \lesssim f_a\lesssim 10^{12}\GEV
$. 
The lower bound  is determined by the stellar cooling arguments for SN1987A~\cite{Mayle:1987as,Raffelt:1987yt,Turner:1987by, Chang:2018rso,Zyla:2020zbs}
and neutron stars (NSs)~\cite{Leinson:2014ioa, Sedrakian:2015krq, Hamaguchi:2018oqw, Beznogov:2018fda, Leinson:2021ety}.
The upper bound is set by the cosmological abundance of the axion.\footnote{The upper bound can be relaxed to be as large as the Planck scale if the Hubble parameter during inflation $H_{\rm inf}$ is lower than the QCD scale and if the inflation lasts long enough~\cite{Graham:2018jyp,Guth:2018hsa}. Similarly, the over-closure problem for the string axion is absent if $H_{\rm inf}\lesssim  0.1$keV~\cite{Ho:2019ayl}. 
}
The lower bound is known to be relaxed for hadrophobic/astrophobic axions  with non-trivial couplings to fermions and gluons~\cite{DiLuzio:2017ogq,Bjorkeroth:2019jtx,DiLuzio:2022tyc}.  However, it is highly non-trivial whether hadrophobic/astrophobic axions with specific PQ charge assignments can be consistent with  GUT. 

Regarding GUT-based axions, they generally possess couplings to both photons and gluons, and consequently, to hadrons. Thus, not only the QCD axion but also axions, in general, must satisfy the limits set by stellar cooling based on nucleon coupling. In this respect, the GUT axion encompasses the parameter region of the QCD axion. The difference between them lies in the origin of their masses, with the general axion obtaining its mass from explicit PQ symmetry breaking other than non-perturbative QCD effects.

In this paper, we demonstrate that axions originating from GUT can be hadrophobic, regardless of whether they are QCD axions or not. Specifically, under certain PQ charge assignments, the axion couplings to mesons and nucleons are suppressed (see also Refs.\,\cite{Bardeen:1986yb,Ema:2016ops,Calibbi:2016hwq} for models with flavor-dependent PQ charge assignment). We propose a novel interpretation of the required PQ charge assignment in terms of the isospin symmetry.\footnote{This condition for charge assignment, along with the interpretation using isospin conservation, was presented in WY's talk (see the detail for https://indico.cern.ch/event/1108846/contributions/4679286/) at "The 2022 Chung-Ang University Beyond the Standard Model Workshop" on  February 9th, 2022.}
Additionally, to satisfy the stringent constraints on the axion-electron coupling arising from the tip of the red giant branch (TRGB)~\cite{Viaux:2013lha,Straniero:2020iyi, Capozzi:2020cbu}, we impose the electrophobic condition on the axion for $f_a$ near the lower end of the axion window. Subsequently, we classify potential charge assignments that satisfy both hadrophobic and electrophobic conditions and discuss flavor models based on the derived PQ charges.

If the axion in question is the QCD axion, the lower end of the axion window is slightly relaxed, while the photon coupling is the usual DFSZ one.
 Notably, this may revive a relatively small $f_a$ slightly outside the conventional axion window, where some indications of extra stellar cooling exist. This region can be thoroughly investigated in future solar axion helioscope, laser-based experiments, and anomalous K meson decay searches. If the axion is the dominant DM component, most of the parameter region within and around the axion window can be probed in future haloscope and lumped element experiments.

Finally, we investigate a scenario where axionic unification of DM and inflation occurs with the hadrophobic and electrophobic GUT axion, as in the so-called ALP miracle scenario~\cite{Daido:2017wwb, Daido:2017tbr}. In this scenario a single axion with an upside-down symmetric potential accounts for both inflation and DM. The original ALP miracle scenario can be fully tested by axion helioscopes~\cite{Irastorza:2011gs, Armengaud:2014gea, Armengaud:2019uso, Abeln:2020ywv,Anastassopoulos:2017kag}, laser-based collider experiments~\cite{Hasebe:2015jxa,Fujii:2010is,Homma:2017cpa,Homma:2021hnl, Homma:2022ktv}, and dark radiation measurements in future cosmic microwave background (CMB)  and baryonic acoustic oscillation (BAO) experiments~\cite{Kogut:2011xw, Abazajian:2016yjj, Baumann:2017lmt}. Until now,  the embedding of the ALP miracle scenario in GUT has not been discussed due to the stringent astrophysical constraints arising from the axion-nucleon coupling.
We identify a possible GUT-inspired model in which reheating proceeds via ALP-$\tau$ lepton or ALP-charm quark interactions. We also find that successful inflation and structure formation requirements suppress the neutron electric dipole moment (EDM). This is because the QCD-induced potential for the ALP is not entirely upside-down symmetric. To ensure both the plateau hilltop for slow-roll inflation and the small DM mass, the minima or maxima of the induced QCD potential must align with the bare potential minima or maxima of the ALP~\footnote{In Ref.\,\cite{Takahashi:2021tff} heavy QCD axion inflation was considered. Here the strong CP phase is aligned to the phase in the inflaton potential due to small QCD instanton effect. It was {discussed in the minimal setup that the parameter region that axion becomes the dominant DM has a too small decay constant to be consistent with the SN1987A and NS bounds}. 
}. Interestingly, the QCD contribution is of a suitable magnitude to suppress the EDM just below or around the current bound, suggesting that this scenario may be probed in the near future.

This paper is organized as follows. In the next section, we review the basics of the hadrophobic axion and the constraints in the context of the effective theory (EFT). In \Sec{HGA} we provide a generic discussion for the hadrophobic GUT axion in  a GUT-inspired EFT. Then we show the parameter region for the generic axion and QCD axion featuring $\SU(5)\times \U(1)_{\rm PQ}$ symmetry.    In Sec.\,\ref{app:1} we build some flavor models resulting the GUT EFT and discuss the flavor structure. In \Sec{mir} we study the scenario that an axion with universal coupling explains both inflation and DM.  The last section is devoted to conclusions. In Appendix \ref{chap:bound} we consider the star cooling bounds and hints, and study them by taking account of the hadronic uncertainty.

\section{Review on hadrophobic axion}
\subsection{Conditions for hadrophobia} 
\label{sec:2}

For clarity, let us first review the QCD Lagrangian with two flavors in the following form:
\beq
\laq{lag}
-{\cal L}=m_u \bar{u}e^{i\gamma_5 c_u \frac{ a}{ \sqrt{2} v_a}} u+ m_d \bar{d}e^{i\gamma_5 c_d \frac{ a}{ \sqrt{2} v_a}} d,
\eeq
where $c_{I}$ is the axion coupling constant to the  fermion $I$ in the broken phase of the electroweak (EW) symmetry. {Here and in what follows $I$ represents the flavor index, and for instance,  in Eq.~(\ref{eq:lag}), $I$ is $u$ or $d$.} For a while, we use $a$ and $v_a$ to denote the axion and the PQ breaking scale. We do not specify whether $a$ is the QCD axion or a generic axion. An important assumption is that in this basis, there is neither gluon coupling nor derivative couplings to $u$ and $d$. {This is equivalent to the condition that heavy quarks do not contribute to the color anomaly.\footnote{This condition can be easily satisfied by adding a certain number of heavy PQ quarks.} One can understand the necessity of this condition by noting that the color anomaly of heavy quarks induces the mixing between the axion and $\eta$, leading to large axion couplings to nucleons.
} In addition to this Lagrangian, couplings to leptons {and} photons are allowed. We will consider these later.

One can see that the coupling between $a$ and a nucleon is suppressed by the factor of $\mathcal{O}(m_u, m_d)/f_\pi$, where $f_\pi \approx 92$ MeV represents the pion decay constant. Nevertheless, a potential mixing between the pion and $a$ can occur, leading to the unsuppressed axion-nucleon coupling. In fact, the potential for the axion-pion system is given by (see e.g.~\cite{GrillidiCortona:2015jxo})
\beq
\laq{pot}
V_a= -B_0 f_\pi^2 \sqrt{m_u^2+m_d^2+2 m_d m_u \cos{\left[(c_u+c_d)\frac{a}{\sqrt{2} v_a}\right]}} \cos[ \frac{\pi_0}{f_\pi} - \phi_a],
\eeq
with
\beq
\tan[\phi_a] \equiv \frac{(m_u \sin[ c_u\frac{ a}{\sqrt{2} v_a}]-m_d \sin[ c_d\frac{ a}{\sqrt{2} v_a}])}{(m_u \sin[ c_u\frac{ a}{\sqrt{2} v_a}]+m_d \sin[ c_d\frac{ a}{\sqrt{2} v_a}])},
\eeq
where  $B_0$ is a parameter for the chiral condensate fixed by the pion mass.
The above potential induces the mixing between $a$ and a neutral pion, $\pi_0$.
Thus, this potential could induce the axion-nucleon coupling, if the mixing is nonzero. Interestingly, one can show that the mixing can be suppressed~\cite{DiLuzio:2017ogq, Bjorkeroth:2019jtx,DiLuzio:2022tyc}, $\tan[\f_a] \simeq 0$, if
\beq
c_u m_u -c_d m_d \simeq 0.
\eeq
This is nothing more than the limit in which the axion-light quark coupling conserves  isospin, since the coupling is then approximately, $m_u c_u \frac{a}{\sqrt{2} v_a} (\bar{u} i\g_5 u+\bar{d} i\g_5 d)$,
 at the leading order of $\O(a/\sqrt{2} v_a)$. Then, in the limit of the isospin symmetry, the mixing between $a$, which is an isospin singlet, and the pion, which is an isospin triplet, is absent. 
As a consequence, $a$ has only a suppressed coupling to nucleon $g_{aN}$, {which is defined by
\beq
-{\cal L} = i g_{aN} \,a \bar{\Psi}_N\gamma_5 \Psi_N,
\eeq
where $\Psi_N$ represents the nucleon field, and $N$ is equal to $n$ or $p$.}
{Since the axion-nucleon coupling  emerges proportionally  to the breaking of the chiral symmetry or isospin symmetry,
  the nucleon coupling is suppressed as}
\beq
g_{aN}= \O(\frac{m_d}{f_\pi}, \frac{c_u m_u-c_d m_d}{c_u m_u +c_d m_d} )   \frac{m_N}{\sqrt{2} v_a}.
\eeq
{Here, we use $f_\pi$ to represent the typical scale of QCD and provide an order-of-magnitude estimate for the first suppression factor due to explicit chiral symmetry breaking. The second term signifies the mixing effect pertinent to the isospin symmetry breaking.}

It is interesting to note that the quantized couplings~\cite{DiLuzio:2017ogq, Bjorkeroth:2019jtx,DiLuzio:2022tyc} \beq \laq{hadrophobic} \boxed{c_u=2/3 c_3,~~~ c_d=1/3 c_3}\eeq 
lead to a good conservation of the isospin
because the light quark masses approximately satisfy the relation, $m_d \simeq 2 m_u$. Here $c_3$ is the anomaly coefficient of the axion-gluon coupling, to be defined in the next subsection. 
 In this case we have $g_{aN}= \O(0.01) \frac{c_3 m_N}{\sqrt{2} v_a}.$ 
  Note that this condition is written under the assumption of two-flavor QCD. Namely, in order to impose this relation for obtaining hadrophobic axion on the theory, we must first integrate all heavier fermions. Then, one can see that $c_3$ can be interpreted as the anomaly coefficient of gluons (as shown in \Eq{derivative}) when we remove the phase factor including $a$ in the quark masses in \Eq{lag}. Thus,   
\beq
c_3= c_u+c_d.
\eeq
In this basis the axion-quark  coupling occurs via the derivative. {In the UV completions we consider in this paper, $c_3$ is an integer so that $a$ has a periodic condition $a\approx a+2\pi (\sqrt{2}v_a)$ representing the phase of some PQ scalar field.}

In the following, we will focus on the charge assignment with $c_3=\mathcal{O}(1)$. Our argument is consistent with Ref.\cite{GrillidiCortona:2015jxo}, which provides the precise values of the axion-nucleon coupling constants, $c_n$ and $c_p$:
\begin{align}
\laq{precise}
c_n\equiv\frac{\sqrt{2} v_a}{m_n} g_{an}= -0.02 \pm 0.03, \quad \text{and} \quad c_p\equiv\frac{\sqrt{2} v_a}{m_p} g_{ap}= -0.02 \pm 0.03.
\end{align}
Here, we take $c_u=2/3$ and $c_d=1/3$ and neglect contributions from heavy flavors. Both constants are consistent with zero.

{It is customary to define the axion-photon coupling $g_{a \gamma \gamma}$ by
\beq
\label{gagg}
\delta {\cal L} = -\frac{1}{4}g_{a \gamma \gamma} a F_{\mu \nu} \tilde{F}^{\mu \nu},
\eeq
where $F_{\mu nu}$ and $\tilde{F}^{\mu \nu}$ are the field strength of photons, and its dual, respectively. 
{In this basis, the} contribution of the axion-pion mixing to the axion-photon coupling is} also suppressed as\footnote{
The same suppressed coupling was obtained from the cancellation in KSVZ-like axion models, which opens the so-called hadronic axion window \cite{Chang:1993gm, Moroi:1998qs}. In our scenario, the axion has a heavier mass than the hadronic axion window in the trapped region for the SN1987A cooling due to the suppressed nucleon couplings.
On the other hand, in some GUT models motivated by the triplet-doublet problem, the axion-gauge/matter coupling relation of GUT can be altered while maintaining hadrophobia. In this case, the hadronic axion window may be opened due to the weak coupling rather than the trapping effect by taking into account the hadronic ambiguity (see Appendix \Sec{bound}). 
Since, in this case, we do not have to ensure the GUT relation in the EFT, we can consider a universal coupling to all the quarks in $10$-plets, while all the leptons do not have the axion coupling.
In this case, the flavor physics constraint and the red giant star cooling bound can be alleviated. By taking certain gluon coupling and photon coupling on a mass basis, we can have hadrophobia and photophobia to open the window. To clarify this possibility, we may need to consider some specific GUT model-building.
}
\beq
\laq{api}
g_{a \g\g}^{a-\pi}= \frac{e^2}{8\pi^2 \sqrt{2} v_a} (8 c_u/3+  2 c_d/3-1.92 (0.04) c_3)\simeq 
 \frac{e^2}{8\pi^2 \sqrt{2} v_a} 0.08 (0.04) c_3.
\eeq
{However, as we will see, one has to include the contribution to the axion-photon coupling that is induced when we switch to the basis given by \Eq{lag}. This reproduces the well-known formula for the axion-photon coupling. }

Before moving on to the next part, let us comment on the axion mass formula. Although the mixing with a pion is absent (or suppressed), the axion does have the usual potential from non-perturbative QCD effects. To see this, let us integrate out the combination $\pi_0/f_\pi-\phi_a$ in \Eq{pot} to obtain the effective QCD potential for $a$, 
\begin{align}
\laq{inst}
V_{\rm QCD}[a]&= -B_0 f_\pi^2 \sqrt{m_u^2+m_d^2+2 m_d m_u \cos{\left[(c_u+c_d)\frac{a}{\sqrt{2} v_a}\right]}}.
\end{align}
This potential only depends on the combination $(c_u+c_d)= c_3$. For later convenience, we define $\chi=\frac{m_u m_d B_0 f_\pi^2}{\sqrt{(m_u+m_d)^2}}\approx (76\MEV)^4$ as the topological susceptibility.

\subsection{EFT description of axion, and some relevant constraints}
We will now describe the hadrophobic axion in the EW symmetric phase. In general, we can consider a low-energy EFT that consists of the SM particles plus  axion up to dimension 5 operators as follows:
\begin{align}
\laq{derivative}
\d \mathcal{L}=&-\frac{a}{\sqrt{2} v_a} \sum_{i=1}^3\frac{c_i g_i^2}{32\pi^2} F_i\tl{F}_i - \frac{\partial_\mu a}{\sqrt{2} v_a} J_{\rm PQ}^\mu.
\end{align}
Here, $i=1,2,3$ denotes the $\U(1)_Y, \SU(2)_L, \SU(3)_c$ gauge groups, respectively, $F_i (\tl{F}_i)$ is the field strength (its dual), and we adopt the GUT normalization $g_1= \sqrt{5/3} g_Y$. Here $c_i$ represents the anomaly coefficient of the axion-gauge coupling. 
{The axion-photon coupling is written as
\beq
g_{a \gamma \gamma}= \left(\frac{5}{3} c_1+c_2{-1.92}\right)\frac{\alpha}{2\sqrt{2} \pi v_a}.
\eeq
}
{Here the third term, $-1.92$, arises when we switch to the basis given by \Eq{lag}~\cite{DiLuzio:2017ogq}.}

The PQ current $J_{\rm PQ}^\mu$ is given by
\begin{align}
J_{\rm PQ}^\mu = \sum_{\alpha}q_\alpha \bar{\psi}_\alpha \bar\s^\mu \psi_\alpha + q_H i (H^* D^\mu H- H D^\mu H^* ),
\end{align}
where $q_\alpha$ is the PQ charge of the SM fermion $\psi_\alpha$ in the chiral representation, $q_H$ is the PQ charge of the Higgs field {$H$}, and $\alpha=Q_{I},u_{R,I},d_{R,I},L_{I},e_{R,I}$.\footnote{{For notational simplicity, we drop ``c" for the right-handed anti-fermions.}} Here, with the abuse of notation,  $I$ represents the flavor index of the corresponding fermions in the chiral representation. For instance, $I=u,c,t$ for up-type quarks, $I=d,s,b$ for down-type quarks, and $I=e,\mu,\tau$ for charged leptons. 
We use the chiral representation of the fermions in the following. {The left-handed quark doublets, $Q_I$, are} defined as
\begin{align}
Q_I\equiv \left(u_{L,I}, (V_{\rm CKM})_{IJ}d_{L,J}\right),
\end{align}
where $V_{\rm CKM}$ is the Cabibbo-Kobayashi-Maskawa (CKM) matrix,
{and $I$ and $J$ run over $u,c,t$ and $d,s,b$, respecitvely.} We assume the absence of flavor mixing in the limit of $V_{\rm CKM}\to 1$ for simplicity, as in the case of minimal flavor-violation (MFV).
In the identity limit of $V_{\rm CKM}\to 1$, 
we can relate {the axion-fermion coupling} $c_I$ and {the PQ charge} $q_\alpha$ by rotating the phase of fermions to move $a$ to the fermion mass terms,
\beq
\laq{rela}
c_I= q_{Q_I}+q_{u_{R,I}}, ~q_{Q_I}+q_{d_{R,I}}~\AND~  q_{L_I}+q_{e_{R,I}}
\eeq
for up-type quark, down-type quark, and charged lepton, respectively. {Here the axion coupling to the fermion $I$ is defined as in \Eq{lag}.}

Neglecting flavor mixing {beyond the MFV}, the Lagrangian in Eq.~(\ref{eq:derivative}) is the most general one obtained by integrating out heavy fields other than the SM particles and $a$. We note that the PQ charge assignment $q_\alpha$ may not allow some of the SM Yukawa interactions, in which case the corresponding Yukawa interactions must arise from the spontaneous PQ breaking (we will discuss some UV renormalizable models in Sec. \ref{app:1}).
The last term in Eq. \eq{derivative} involves a derivative of $a$ and satisfies the shift symmetry, $a\to a+C$, with $C$ being an arbitrary real number. Thus, this term neither generates an axion-gauge boson Chern-Simons coupling nor the mass term of $a$. The anomaly matching in this basis should be solely satisfied by the first term.

In the following, we take \beq q_H=0\eeq by a field redefinition and a redefinition of $q_\a.$\footnote{This process does not violate the following GUT relation of 
the PQ charge for fermions if it is  understood as the process acting on the $5_H$. } 
In this basis, $a$ is not eaten by the $Z$-boson when the Higgs field gets a vacuum expectation value (VEV).
We will not further reduce the Lagrangian redundancy by field redefinitions for our later purpose.

A stringent bound on this EFT comes from flavor violation via the CKM mixing. 
The left-handed quark part in the broken phase is given by~\cite{Feng:1997tn}
\beq
J^\mu_{\rm PQ}\supset \sum_{I} q_{Q_I} \bar{u}_{L,I} \g^\m u_{L,I }
+\sum_{I,J} \bar{d}_{L, I } (q_{Q, \rm CKM})_{IJ} \gamma^\m d_{L, J} 
\eeq
with
\beq
(q_{Q, \rm CKM})_{IJ}\equiv \sum_{K}{ (V^*_{\rm CKM})_{KI}q_{Q_K} (V_{\rm CKM})_{KJ}}.
\eeq
In particular, the most severe bound for our setup comes from the $K^+\to \pi^+ a$ process: $Br_{K^+\to \pi^+ a}<5\times 10^{-11} ~(90\% \,\rm CL)$  \cite{NA62:2021zjw} (See also \cite{E949:2008btt}). Following Refs.~\cite{Goldman:1977en, MartinCamalich:2020dfe }, we obtain
\beq
\laq{Kpia0}
\frac{\(q_{Q, \rm CKM}\)_{ds}}{2 \sqrt{2} v_a}< 1.2\times 10^{-12} \GEV^{-1}. 
\eeq

For $\sqrt{2} v_a\ll 10^{12}\GEV$, 
we only have the possibility of
 $q_{Q_u}=q_{Q_c}\equiv q_{Q_{u,c}},$ as in the 
 Glashow-Iliopoulos-Maiani mechanism.  
Then we get 
$|(q_{Q, \rm CKM})_{d s}|\approx 3.5\times10^{-4}|q_{Q_{u,c}}-q_{Q_t}|$, and
arrive at the lower bound on $v_a$,
\beq
\laq{Kpia}
\sqrt{2} v_a \gtrsim 3.0\times 10^8\GEV |q_{Q_{u,c}}-q_{Q_t}|.
\eeq
{The future reach, $Br_{K\to \pi a}\lesssim 10^{-11}$, in NA62 and KOTO experiments \cite{Ruggiero:2017hjh, Nanjo:2019chy,MartinCamalich:2020dfe}} corresponds to
\beq \sqrt{2} v_a \lesssim 7 \times 10^8\GEV  |q_{Q_{u,c}}-q_{Q_t}|.\eeq 
The other flavor constraints are  weaker~\cite{MartinCamalich:2020dfe}.

For  later convenience, let us also write down several bounds on other fermion couplings. 
A stringent bound on the axion-electron coupling comes from the tip of the red giant branch (TRGB)~\cite{Straniero:2020iyi}(see also \cite{Viaux:2013lha, Capozzi:2020cbu}),
\beq
\laq{TRGB}
g_{aee}\equiv \frac{c_e m_e}{\sqrt{2} v_a}<  1.48\times 10^{-13} ~~[95\% \text{ CL} ],
\eeq
which is valid for $m_a\ll 10\KEV$.
For $c_e=\O(1)$, we thus have $\sqrt{2} v_a\gtrsim 10^{9}\GEV$. In fact, 
there is also a hint for extra cooling~\cite{Straniero:2020iyi}(see also \cite{Viaux:2013lha, Capozzi:2020cbu}), which corresponds to
\beq
\laq{TRGhint}
g_{aee}= 0.6^{+0.32}_{-0.58}\times 10^{-13}.
\eeq

The TRGB also imposes a bound on the axion-top coupling via  loop effects, even with $c_e=0$ in the UV. 
If the axion-top coupling is non-zero, we need to carefully consider the radiative corrections within the EFT. 
The axion-top coupling induces an effective PQ charge on the Higgs field at one-loop level~\cite{Feng:1997tn} in the symmetric phase (see also Refs.~\cite{Choi:2017gpf, Chala:2020wvs, Bauer:2020jbp, Chakraborty:2021wda, Bonilla:2021ufe}). The corresponding Lagrangian is given by
\beq
\d {\cal L}_{\rm eff}= i \d q_H (H^* D_\mu H- H D_\mu H^*) \frac{\partial^\mu a}{\sqrt{2} v_a},
\eeq
where
$\d q_H\sim \frac{3 y_t^2 |c_t|}{16\pi^2}\log{(\L_{\rm cutoff}^2/\mu^2_{\rm RG})}$ with $\L_{\rm cutoff}$ being the cutoff scale below which the EFT of the SM plus $a$ is valid, 
and $\mu_{\rm RG}$ is the renormalization scale. 
We need to perform a chiral rotation to remove $a$ from the phase in the Higgs field so that the $Z$-$a$ mixing is absent after the EW symmetry breaking. As a result, the axion
$a$ acquires an additional suppressed couplings to all of the SM fermions. 
In particular, the axion-electron coupling with the boundary condition $c_e[\L]=0$ is obtained as
\beq
g_{aee}\simeq (3-10)\times 10^{-13} |c_t| \frac{10^8\GEV}{\sqrt{2}v_a}
\eeq
by varying $\L_{\rm cutoff}=(1-10^2)\TEV$ and $\mu_{\rm RG}=100\GEV$.\footnote{
Usually, this bound is omitted in previous studies by taking $\m_{\rm RG}=\L_{\rm cutoff}$. 
In this case, however, the collider constraints and the threshold effect by integrating out the heavy field become important. 
}  
This is in tension with the TRGB bound \eq{TRGB} if \beq \sqrt{2} v_a/|c_t| \lesssim (2\,\text{-}\,6)\times 10^8\GEV.\eeq

\section{Hadrophobic axion from GUT}
\lac{HGA}

Let us have a general discussion based on $\SU(5)\times \U(1)_{\rm PQ}$.
Below the PQ breaking scale we can switch to the basis where $a$ is absent in the phase of the fermion masses, and then integrate out the heavy degrees of freedom other than the SM particles and $a$. 
In the low energy EFT with the SM plus $a$, the GUT relation naturally resides in the axion derivative couplings. 
(We will explicitly discuss two concrete UV models in Sec.\,\ref{app:1}).  
In $\SU(5)$ GUT, we have
$
{10}_x= \{ Q_x, u_{R,x},e_{R, x}\}$, ${ \bar{5}}_x= \{L_x,d_{R, x} \} $ with $x=1,2,3$ being the index of the GUT generation. 
The GUT embeddings are expressed as, 
\beq
\laq{qGUT}
q_{u_{R,I}}=q_{Q_{J}}=q_{e_{R,K}}=q_{10_x},~~ q_{d_{R,I}}=q_{L_{J}}=q_{\bar{5}_x}, \AND
c_1=c_2=c_3=c_{5}. 
\eeq
Here we include the possibility that the embedding may not be flavor-blind i.e. $I,J,K$ may be {in different generations}, and {for satisfying the MFV assumption, {we consider} that $x$ is aligned with $I,J,K$.  {In other words, each SM multiplet enters once in the corresponding GUT multiplet, and no linear combination such as $Q_{u} + Q_{c}$ appears.\footnote{{Namely this is the requirement that the GUT gauge interaction satisfy the MFV. 
 On the other hand, another embedding via MFV is to multiply $V_{\rm CKM}$ accordingly to the up-type or down-type right-handed quarks. For the purpose to show the possible origin of the hadrophobic axion, we do not consider this possibility.
}}  The universal condition on the axion-gauge coupling comes from the 't Hooft anomaly matching and \Eq{sym}.}

Before going into details, 
{we remind}
the general prediction for {the} axion from GUT. 
{Although} the hadronic contribution to the photon coupling, \eq{api}, is {negligibly small} {in the basis we adopted}, the
photon and gluon couplings via \Eq{qGUT}, and the chiral anomaly for changing the basis {reproduce} the usual GUT axion-photon coupling
\beq
\laq{gagg}
{g_{a\g\g} \simeq 0.87\times 10^{-11}\GEV^{-1} \(\frac{10^8\GEV}{\sqrt{2} v_a/c_3}\)}.
\eeq
{See Refs.~\cite{Kim:1979if, Shifman:1979if,Dine:1981rt,Zhitnitsky:1980tq} for the models of KSVZ and DFSZ axions.}

In the rest of this paper, we focus on the case where $\bar{5}_x$ has a flavor-blind PQ charge,
\beq
q_{\bar{5}_x}=q_{\bar5},
\eeq
 so that the PQ symmetry allows large neutrino mixing. In the following, we discuss the model-building for $v_a\gtrsim10^{9}\GEV$ and $v_a\lesssim 10^{9}\GEV$ separately, considering the limits of the TRGB bound. At the end of this section we give the viable parameter range for the axion DM in our proposal.

\subsection{Models for $v_a\gtrsim10^{9}\GEV$}

The most simple PQ charge assignment for $10_x$ is the flavor-blind one,
\beq
q_{10_x}=q_{10}. 
\eeq
 From \Eqs{hadrophobic}, \eq{rela} and \eq{qGUT},
we can find the hadrophobic condition:
\beq
\boxed{\text{Flavor-blind realization}, v_a\gtrsim 10^{9}\GEV: q_{10}=c_{5}/3,~~ q_{\bar{5}}=0.}
\eeq
In other words, once this charge assignment is given in the GUT theory, the hadrophobic condition is accidentally satisfied.
Note that we should include additional PQ charged fermions to give the corresponding gauge anomaly in this case.
We emphasize that this charge assignment predicts $c_e=c_{5}/3$ and $v_a/c_e$ must be larger than
$\sim 10^{9}\GEV $ to satisfy the  TRGB bound~\eq{TRGB}.
Such flavor-blind model-building is relatively easy, see e.g. Refs.~\cite{Takahashi:2019qmh,Takahashi:2020bpq}.

When $v_a \gtrsim 10^{11}\GEV$, the $K^+ \to \pi^+ a$ bound is no longer important for any flavor-dependent PQ charge assignment. The model can have arbitrary quark flavor structure. For instance, one can have $q_{10_1}=c_5/3$ while the others are $0$. Here $10_1$ involves $Q_u$ and $u_R.$ A similar model can be found in Ref.\,~\cite{Bardeen:1986yb}, which does not suffer from the cosmological domain wall problem.

{It is also worth noting that one {does not need to} introduce a singlet scalar field to  build the hadrophobic axion when $v_a\gg 10^{11}\,\GEV$. One can introduce a {complex} adjoint Higgs field responsible for the  breaking of GUT~\cite{Wise:1981ry,FileviezPerez:2019ssf}, which also breaks the PQ symmetry. This is one of the simplest models. The Yukawa coupling structure can easily be compatible with the charge assignment by introducing higher dimensional terms e.g.~\Eq{hg}. }

\subsection{Models for $v_a\lesssim 10^{9}\GEV$.}
\lac{flavor}
Now we turn our attention to the case of $v_a\lesssim 10^{9}\GEV$, where we have to be careful about the limits of flavor violation~\eq{Kpia0} and TRGB~\eq{TRGB}. {Indeed, in this regime we need the axion also to be electrophobic as well.}
Without loss of generality, we can define $10_1\supset Q_{u}, 10_2 \supset Q_{c}$. Then, we need
\beq q_{10_1}=q_{10_{2}}\equiv q_{10_{1,2}}\eeq
to satisfy \Eq{Kpia0} or \Eq{Kpia}. 
A possible way to satisfy \Eq{TRGB} in this case is to postulate that one of the $10$s, denoted by $10_x$,  satisfies
\beq \laq{ce}
c_e=q_{10_{x}}+q_{\bar{5}}=0,
\eeq  
where $10_x$ hosts $e_R$. Alternatively, we can simply suppress the axion electron coupling
by e.g. the mixing effect in the (extended) Higgs sector~\cite{DiLuzio:2017ogq, Bjorkeroth:2019jtx,DiLuzio:2022tyc}, which we do not consider in this paper. 
In fact, the possibility of $x=1,2$ in \Eq{ce} is excluded since it would imply $c_d = 0$ and  we are interested in the case of $c_u,c_d,c_{3}\neq 0$.
So we have
$x=3 $, i.e., the right-handed electron must be embedded in $10_3$. 

In addition, the hadrophobic axion requires 
\beq
\laq{had2}
q_{10_{1,2}}+q_{10_{y}}= 2(q_{10_{1,2}} +q_{\bar{5}}) ,~~~c_{5}= 2q_{10_{1,2}}+q_{10_{y}}+ q_{\bar{5}},
\eeq
where $10_y$ hosts $u_R$, and the first and second conditions correspond to $c_u=2c_d$ and  $c_3= c_u + c_d$, respectively.  Here we introduce $10_y$ because we do not specify where $u_R$ is embedded at this point.

The above conditions 
leave us with only two possible charge assignments. 
The first possibility corresponds to the choice of $y=1$ or $2$ in \Eq{had2}, i.e., both left and right-handed up-type quarks are in $10_1 \OR 10_2$. In this case, 
 we must have $q_{\bar{5}}=0$ from \Eq{had2}. This leads to $q_{10_3}=0$ from \Eq{ce}. Thus we obtain\footnote{
 The PQ charges assigned for $10$ and $\bar{5}$ fermions can taken to be both flavor-blind if 
the axion mass is higher than the typical temperature of the red giant stars $\sim \O(10)\KEV.$ 
For hadrophobia, we need \Eq{hadrophobic} in the effective theory by integrating out the other fermions. This condition can be easily satisfied along with the flavor-blind charge assignment if we introduce certain additional heavy fermions. 
In this case the axion emission rate is suppressed in the red-giant.
Then the stringent TRGB bound is evaded (and perhaps further explains the TRGB hint). 
 } 
\beq  \boxed{\text{\it model 1: } q_{10_{1,2}}=c_{5}/3 ,  ~~~q_{10_3}= q_{\bar 5}=0. }\laq{rel} \eeq
Since $Q_u, \AND Q_c$ are embedded in $10_{1,2}$, $Q_t$ is in $10_3.$ 
Since $e_R$ is in $10_3$ to satisfy the stellar cooling bound, $\mu_R, \tau_R$ are in $10_{1,2}$. Then the PQ charges are given by $q_{Q_{u},Q_{c},{\m}_R,\t_R, u_R}= c_{5}/3$ and $q_{Q_{t}, e_R, L_e, L_\mu, L_\tau, d_R, s_R, b_R}=0$.\footnote{We have used the short hand notation, $e_{R,I}\to I_{R}, d_{R,I}\to I_{R}.$ For instance, $e_{R,e} \to e_R$.}
We have two kinds of embedding of $t_R$ and $c_R$:  normal embedding, $\{10_{1,2}, 10_3 \}\supset \{c_R,t_R\}$ and  inverted embedding, $\{10_{1,2}, 10_3 \}\supset\{t_R, c_R\},$ 
 which give $\{q_{c_R},q_{t_R}\}=\{c_{5}/3, 0\}, \AND \{0, c_{5}/3\}$, respectively. 
Namely the charge assignments are given as,
\begin{align}
\text{{\it model 1} - Normal}: q_{Q_{u},Q_{c},{\m}_R,\t_R, u_R, c_R}= c_{5}/3 \AND q_{Q_{t}, e_R, L_e, L_\mu, L_\tau, d_R, s_R, b_R, t_R}=0, \\
\text{{\it model 1} - Inverted}:  q_{Q_{u},Q_{c},{\m}_R,\t_R, u_R, t_R}= c_{5}/3 \AND q_{Q_{t}, e_R, L_e, L_\mu, L_\tau, d_R, s_R, b_R, c_R}=0.
\end{align}
Therefore we arrive at two types of $c_I$ prediction as
\begin{center}
  \begin{tabular}{@{} ccccccccccccc @{}}
    \hline
{\it model 1}&  $c_u$ &$c_c$&$c_t$&$c_d$&$ c_s$&$c_b$ &$c_e$&$c_\m$&$c_\t$\\ 
\hline Normal &$2c_{5}/3$ & $2c_{5}/3$ & 0 & $c_{5}/3$ & $c_{5}/3$ & 0 & 0& $c_{5}/3$ & $c_{5}/3$ & \\ 
    \hline Inverted &$2c_{5}/3$& $c_{5}/3$ & $c_{5}/3$ & $c_{5}/3$ & $c_{5}/3$ & 0 & 0& $c_{5}/3$ &$c_{5}/3$ \\ 
    \hline
  \end{tabular}
\end{center}
~\\

If $y=3$ in \Eq{had2}, on the other hand, we obtain $q_{10_{1,2}}+q_{\bar 5}=c_5/3.$ Since $c_5$ should not be zero for our purposes, electron should be in $10_3$, i.e., $x=3$, {which is consistent with what we  found earlier.}
So we get
\beq  \boxed{\text{\it model 2: } q_{10_{1,2}}=c_{5}/2,  q_{10_3}=-q_{\bar 5}=c_{5}/6 . }\laq{rel2} \eeq
Since $u_{R}$ is in $10_3$,  $c_R, t_R$ are in $10_{1,2}$.
Following a similar argument, we get the charge assignment,
\beq \text{\it model 2:~} q_{Q_{u},Q_{c},{\m}_R,\t_R, c_R, t_R}= c_{5}/2,q_{Q_{t}, u_R, e_R}=c_{5}/6 , \AND  q_{L_e, L_\mu, L_\tau, d_R, s_R, b_R}=-c_{5}/6.\eeq 
Accordingly, $c_I$ is given by 
\begin{center}
  \begin{tabular}{@{} ccccccccccccc @{}}
    \hline
{\it model 2} &  $c_u$ &$c_c$&$c_t$&$c_d$&$ c_s$&$c_b$ &$c_e$&$c_\m$&$c_\t$\\ 
\hline  &$2c_{5}/3$ & $c_{5}$ & $2c_{5}/3$ & $c_{5}/3$ & $c_{5}/3$ & 0 & 0& $c_{5}/3$ & $c_{5}/3$ & \\ 
    \hline
  \end{tabular}
\end{center}

In both models, we have $|q_{Q_t}-q_{Q_{u,c}}|= c_{5}/3$.
This gives $\sqrt{2}v_a\gtrsim 10^{8}\GEV $ from \Eq{Kpia} for $c_{5}=\O(1)$.
This is comparable to the astrophysical bounds of SN1987A, NSs, and horizontal branch (HB) stars as will be discussed.

\paragraph{SM fermion mass hierarchy}
Roughly speaking, when $c_I\neq 0$, the corresponding Yukawa coupling can be more or less suppressed depending on the model building. This is because, if $q_I$ are considered as the charge assignment in a concrete UV PQ symmetry (in this case, the assumption of $q_H=0$ is a constraint for the model), the non-vanishing $c_I$ forbids the corresponding Yukawa coupling, and the SM Yukawa coupling should be generated via the spontaneous PQ symmetry breaking. We will explicitly show this kind of suppression in Sec.,\ref{app:1}. In this sense, the most natural embedding may be the normal embedding of {\it model 1}, since the top Yukawa coupling is not suppressed. Interestingly, the quarks lighter than $b,t$ naturally have suppressed masses. The most unnatural point in this model is the electron mass, which is much smaller than the bottom mass. This can be explained by an additional symmetry.\footnote{Alternatively, we can consider anthropic selection~\cite{Agrawal:1997gf} through the corrections when we integrate out the heavy UV physics relevant to spontaneous GUT breaking. An example in a different context can be found in~\cite{Endo:2019bcj}, where the threshold corrections of the supersymmetric (SUSY) partners suppress the electron mass and enhance the electron $g-2$. In our case, the small bottom Yukawa coupling may also be related to the anthropically small electron mass via the quark-lepton mass relation.}

In the following, we will concentrate mainly on the normal embedding of the {\it model 1} for {the natural order of the Yukawa couplings}, but we will also comment on the other cases.

\subsection{Parameter region of hadrophobic GUT axion} 

Let us estimate the viable parameter region of the hadrophobic GUT axion. 
One of the most interesting cases is that there is no explicit PQ breaking term other than the anomalous coupling to the SM gauge bosons. Then, the axion is the QCD axion whose decay constant $f_a$ is defined by 
\beq
f_a =\sqrt{2} v_a/c_3.
\eeq 
The  mass of the QCD axion is given by 
\beq
\laq{QCDaxionmass}
m_a^2= \frac{\chi}{f_a^2}.  
\eeq
The hadrophobic (and electrophobic) QCD axion window is larger than the conventional one~\cite{DiLuzio:2017ogq,Bjorkeroth:2019jtx,DiLuzio:2022tyc},
\beq
10^{7}\GEV \lesssim f_a\lesssim 10^{12}\GEV. 
\eeq
The lower bound on the decay constant
is set by the stellar cooling argument of 
NSs~\cite{Leinson:2014ioa, Sedrakian:2015krq, Hamaguchi:2018oqw, Beznogov:2018fda, Leinson:2021ety} or
the SN1987A~\cite{Mayle:1987as,Raffelt:1987yt,Turner:1987by, Chang:2018rso,Zyla:2020zbs}, and it is relaxed due to the suppressed axion-nucleon coupling. 
(See also the appendix.\,\ref{chap:bound} for the constraints we adopt and for a way to further relax the lower bound by taking account of the hadronic uncertainty.) {The K meson decay bound{, \eq{Kpia0}, which is for $90\%$ C.L.}, should be comparable to them {with a similar confidence level}. } 
The upper bound is set by the cosmological overproduction of $a$ when the inflationary Hubble parameter is higher than the QCD scale~\cite{Preskill:1982cy,Abbott:1982af,Dine:1982ah}. In fact, it can be relaxed to be as large as the Planck scale when the Hubble parameter is smaller than the QCD scale and if the inflation lasts long enough~\cite{Graham:2018jyp,Guth:2018hsa}. Thus we do not exclude $f_a\gtrsim 10^{12}\GEV$ in the following.

We show the parameter region of the hadrophobic GUT axion, including the QCD axion (black solid line), in the $m_a-g_{a\g\g}$ plane in Fig.~\ref{fig:1}. The prediction for the hadrophobic GUT QCD axion is obtained from \Eqs{gagg} and \eq{QCDaxionmass}. {We also show the prediction of the KSVZ (gray dot-dashed line).}
In the right frame, we show the range of $g_{a\g\g}$ that can be realized with {\it model 1} or {\it model 2}, as well as the range where the charge assignment has more freedom, including the flavor-blind case, $q_{10_x}=q_{10}$. If we would like to have a hadrophobic GUT axion with a mass heavier than the QCD axion by introducing another potential term, we need to fine-tune the strong CP phase to explain the small nucleon EDM. In \Sec{mir}, we give some examples where the small strong CP phase is explained by the requirement of the inflaton DM.
Similarly, in order to have the axion mass lighter than the QCD axion, we need cancellation between contributions from QCD and other effects.\footnote{When we consider axions lighter than the QCD axion, the tuning of the strong CP phase is difficult to explain without changing our prediction. This is because the QCD axion, in this case, is the combination of the two axions coupled to gluons. Then, by integrating out the QCD axion, the lighter (non-QCD) axion does not have the gluon coupling. So the prediction will be changed. In the case of the axion heavier than the QCD axion line, we can introduce an additional QCD axion to solve the strong CP problem.}

\begin{figure}[!t]
\begin{center}  
\includegraphics[width=145mm]{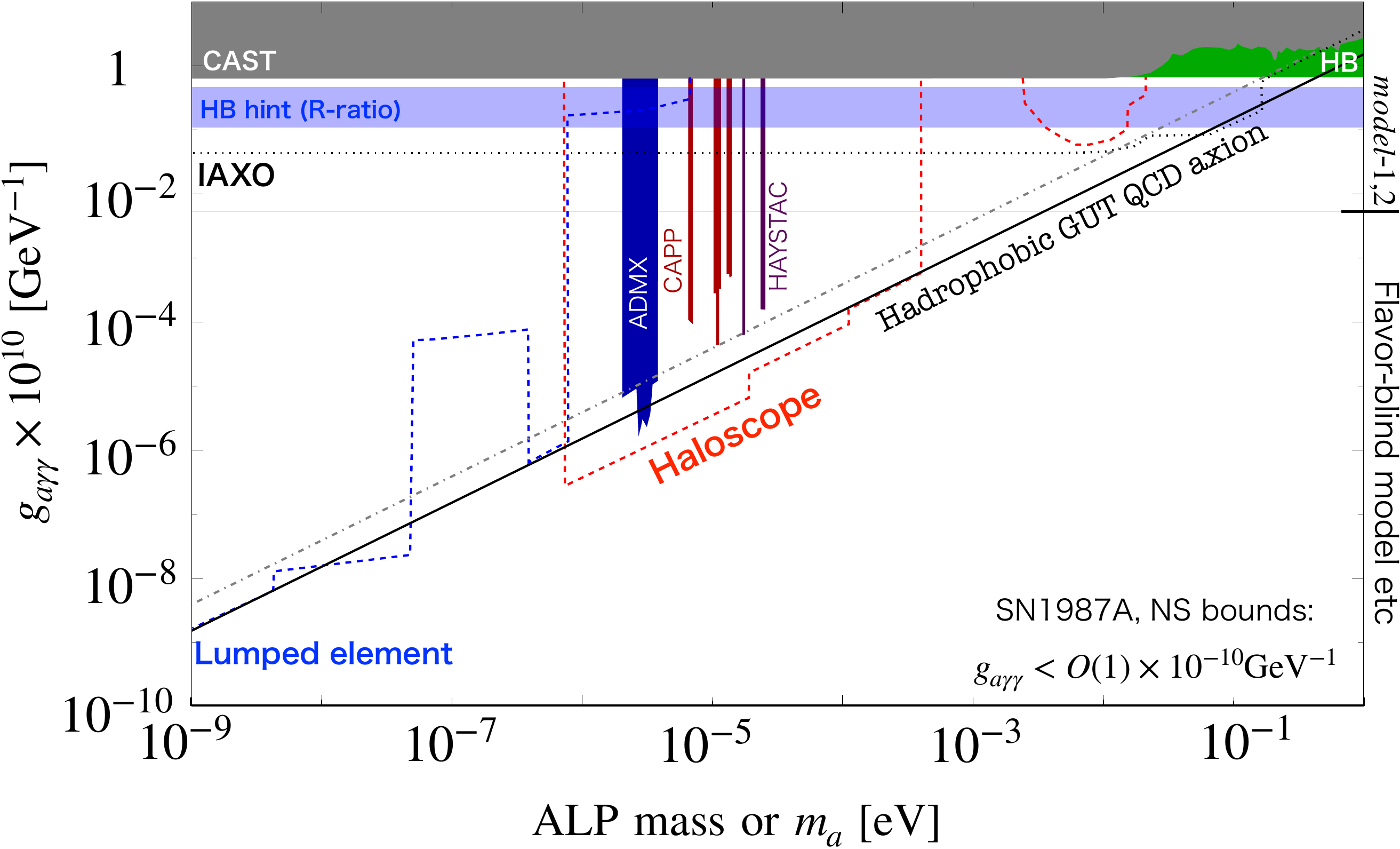}     
      \end{center}
\caption{
The parameter region of the hadrophobic GUT axion (black solid line) in the $m_a -g_{a\g\g}$ plane. 
The KSVZ  axion is denoted by the gray dot-dashed line, the black solid line below which is the hadrophobic GUT axion (the usual DFSZ axion). 
The bounds from solar axion observation by CAST (gray shaded region), HB (green shaded region) are shown. 
Also shown are the future reaches of the IAXO ({black thin dotted line}). 
The cooling hint from HB is in the horizontal blue range. 
The haloscope bounds from ADMX (blue shaded region), CAPP (red shaded region), and HAYSTAC (purple shaded region) experiments assume that the axion composes the dominant component of DM.
The future sensitivity region of the lumped element and haloscope experiments are above the blue and red dashed lines, respectively. 
}\label{fig:1} 
\end{figure}

In the figure, we have included the bounds from the solar axion observation by CAST (gray shaded region) \cite{Anastassopoulos:2017ftl}, the HB cooling (green shaded region) \cite{Raffelt:1985nk,Raffelt:1987yu,Raffelt:1996wa, Ayala:2014pea, Straniero:2015nvc, Giannotti:2015kwo, Carenza:2020zil}, and current DM haloscope searches of ADMX (blue shaded region) \cite{ADMX:1998pbl}, CAPP (red shaded region) \cite{Semertzidis:2019gkj, Lee:2020cfj} and HAYSTAC (purple shaded region) \cite{Brubaker:2016ktl}. 
The future reaches of IAXO (black dashed line)~\cite{Irastorza:2011gs, Armengaud:2014gea, Armengaud:2019uso, Abeln:2020ywv}, 
lumped element experiments (blue dashed line), and haloscope experiments (red dashed line) are also shown (see e.g. Refs.~\cite{Silva-Feaver:2016qhh, Gramolin:2020ict,DMRadio:2022pkf,TheMADMAXWorkingGroup:2016hpc, Brun:2019lyf, Marsh:2018dlj,Schutte-Engel:2021bqm}). 
The reaches and bounds of the lumped element experiments and haloscopes are adopted from Refs.\,\cite{Semertzidis:2021rxs}. We note that the axion DM search experiments assume the axion is the dominant component of DM. 
Interestingly, our QCD axion model can be tested in most of the parameter regions.

We  also plot the hint for the extra cooling based on the $R$-parameter as the horizontal blue region (see \Eq{HB} in the appendix), where we neglect the top-induced electron coupling, which is justified in the normal embedding of the {\it model 1}.
Interestingly, the hint and our prediction overlap in the region that can be probed by the IAXO and TASTE~\cite{Anastassopoulos:2017kag}. 
Moreover, if we consider the inverted embedding of {\it model 1} or {\it model 2}, the TRGB hint (see \Eq{TRGhint}) can also be explained with $\L_{\rm cutoff}=10-100 \TEV$ in this range.  
The cooling hint for the Cassiopeia A neutron star~\cite{Leinson:2014ioa, Hong:2020bxo} also overlaps, although this hint may not be supported by the subsequent analysis~\cite{Hamaguchi:2018oqw,Leinson:2021ety}. 
{The region is slightly in tension with the $K^+\to \pi^+ a$ decay bound, but this can be alleviated by introducing a flavor violating coupling and mild tuning (see Fig.~\ref{fig:ALP} and the explanation in the main text). }
In any case, this region can change the cooling of different stars. So our axion 
 can be probed by further observations of the NSs and by a better understanding of the cooling mechanism. 
In addition, a large part of the parameter space can be probed in laser-based collider experiments \cite{Hasebe:2015jxa,Fujii:2010is,Homma:2017cpa, Homma:2021hnl},
and the aforementioned $K^+\to \pi^+a$ in the NA62~\cite{Ruggiero:2017hjh} and KOTO~\cite{Nanjo:2019chy} experiments in the future. For the latter, see Fig.~\ref{fig:ALP}.

The QCD axion in this interesting region with $g_{a\g\g}\sim 10^{-10}\GEV^{-1}$ has a relatively heavy mass for the QCD axion, and
one has to increase the axion abundance to explain all DM.
Although the original misalignment mechanism~\cite{Preskill:1982cy,Abbott:1982af,Dine:1982ah} suffers from a serious isocurvature problem {when the anharmonic effect is strong}~\cite{Lyth:1991ub,Kobayashi:2013nva}, 
the axion may be produced from the inflaton decay whose reheating temperature is higher than the inflaton mass~\cite{Moroi:2020has, Moroi:2020bkq}.
In this case, the Bose-enhancement effect is under control in the narrow resonance region.

We note that in realistic models we may have a non-trivial domain wall number. If we consider post-inflationary PQ-breaking models,    we need to break the PQ symmetry slightly to let the string domain wall system collapse. Avoiding the quality problem of PQ symmetry (by assuming some discrete symmetry) without tuning,  it has been discussed that there is an overproduction problem of axion DM in the DFSZ model unless the decay constant is comparable to or slightly below the ordinary SN1987A bound (e.g., \cite{Beyer:2022ywc}). In our scenario, however, the decay constant can be smaller than usual, which means that a larger bias can be consistent with the required quality of the PQ quality, and a smaller domain wall stress is also allowed. Thus, we can successfully have the mechanism of DM production from domain wall annihilation.

 \section{Flavor models for hadrophobic and electrophobic GUT axion}
\label{app:1}
In this section, we present two specific models to show that the non-trivial PQ charge assignment required for $v_a\lesssim 10^9\GEV$ in the previous section can be realized.  We focus on the normal embedding of {\it model 1}.\footnote{{The inverted embedding does not change the charge assignment of the fermions but may require different or additional field contents and global flavor symmetries for the Yukawa coupling hierarchy. For instance, one can notice that in the inverted hierarchy case, the assignment of Sec.\,\ref{tabPQ3H} does not allow the top Yukawa coupling. We can introduce another Higgs field $5_{H_{3}}$ that has PQ charge of $-c_5/3$ (the opposite sign of $5_{H_1}$, to write down $5_{H_3} 10_3 10_{1,2}$ which gives top Yukawa coupling. We need to suppress the off-diagonal quark Yukawa matrix such as $5_{H_0} 10_3 10_3$. This can be realized by introducing a certain flavor symmetry.)}}   We show in both UV models that the first two generation quark masses are naturally suppressed. 

The flavor-blind charge assignment, which is useful for $v_a\gtrsim 10^9\GEV$, is straightforward based on the discussions in this section. We will not discuss it explicitly. However, we would like to emphasize that the PQ breaking effect must be large enough to generate the top Yukawa coupling of $\O(1)$, because for the flavor-blind charge assignment, the top Yukawa coupling is forbidden in the PQ symmetric phase. In this case, the ordinary GUT embedding of the quarks and leptons can be fine.

\subsection{A three Higgs doublet model}
A UV realization can be obtained from three Higgs doublet models.\footnote{
{A two Higgs doublet model requires fine-tuning of the ratio of the VEVs usually denoted by $\tan\beta$~\cite{DiLuzio:2017ogq}. This spoils the charge assignment discussion we developed so far. Thus we do not consider the possibility.}
The precision of the gauge coupling unification gets better in the presence of additional Higgs doublets below the intermediate scale. 
}
Let us consider a Lagrangian  with three Higgs doublets, $H_{0,1,2}$ and a PQ scalar field, $\F_{\rm PQ}$, and their PQ charges are given by
\begin{center}
\label{tabPQ3H}
  \begin{tabular}{@{} cccccccccc @{}}
    \hline
Fields &  $\F_{\rm PQ}$&$5_{H_{0}}$&$5_{H_1}$&$5_{H_2}$&$10_{1,2}$&$ 10_3$&$ \bar{5}_{1,2,3}$ \\ 
    \hline
    PQ charge$\times 3/c_{5}$ & 1 & 0& 1 & -2 & 1 & 0 & 0  \\ 
    \hline
  \end{tabular}
\end{center}
Here, $10_{x}$ and $\bar{5}_x$ have the PQ charge assignments of the normal embedding in {\it model 1}. The three Higgs fields are introduced to have Yukawa couplings for all the fermions. 
{Note that the charge assignment is determined by the hadrophobic and electrophobic conditions in the previous section, and this does not guarantee that the above charge assignment correctly reproduces the anomaly coefficient $c_5$. As we will see later, the anomaly coefficient would be different if we assume the above matter content, and we need to add additional matter fields to correctly reproduce the anomalous couplings to the SM gauge bosons.}

{A part of the Higgs potential including the PQ scalar is given by } 
\beq\laq{mix}
{V}\supset
\left(A \F_{\rm PQ}   5_{H_{0}} 5^*_{H_1} + \l (\F_{\rm PQ})^25_{H_2} 5^*_{H_{0}} +h.c.\right) -m_\F^2 |\Phi_{\rm PQ}|^2 + \frac{\l_\F}{2} |\F_{\rm PQ}|^4,
\eeq
where $A,\l,\l_\F$ are coupling constants, and $m_\F^2$ is the mass squared of  $\F_{\rm PQ}.$
In addition, we can also have some portal couplings, e.g. $|H_2|^2 |\F_{\rm PQ}|^2$.
Here we assume they are small enough to simplify the following discussion. {In addition to the above potential, we assume that the Higgs fields have the ordinary potential allowed by the symmetry.}

The Yukawa couplings are given by
\begin{align}
-{\cal L}_{\rm Yukawa} &= \sum_{x,y\neq 3}Y_{10; xy} 5_{H_2} 10_{x}10_y+Y_{5; xy}5_{H_1}^* \bar5_x 10_y\\&+ \sum_{x\neq 3}Y_{5; 3x}5_{H_1}^* \bar5_3 10_x +Y_{5; x3}5^*_{H_0} \bar{5}_x 10_3
\\& +
(Y_{10; 33} 5_{H_{0}} 10_3 10_3+Y_{5; 33} 5^*_{H_{0}} \bar5_3 10_3) +h.c.
\end{align}

Let us assume that $m^2_{\F} (>0)$ is much larger than the EW scale.
Then, the PQ symmetry breaking is triggered when $\F_{\rm PQ}$ develops a nonzero VEV.
We assume  that the  the colored Higgs has a mass around the GUT scale $\sim 10^{15}\GEV$.\footnote{Solving the triplet-doublet splitting problem is beyond the scope of this paper.} Below the GUT scale, 
we have the PQ breaking at the scale of $m_\F,$ and $\vev{\F_{\rm PQ}}=v_a\approx \sqrt{m_\F^2/\l_\F}.$ 
Then we get the mixing term between $H_{0}$ and $H_{1,2}$. 
Suppose that $H_{\rm 1, 2}$ have a positive mass squared $m_{H}^2$, which satisfies $m_H^2 \gg v_{\rm EW}^2, A v_{\rm EW}$ and  $m_H^2\lesssim \O(m^2_\F)$, 
where $v_{\rm EW}\sim 170\GEV$ is the VEV of the SM-like Higgs.  
{The inequality implies that the SM-like Higgs boson mass is obtained by fine-tuning, on which we will comment later.}
Then we can safely integrate out the radial mode of the PQ Higgs boson, $s$, which is defined by the expansion,  $\F_{\rm PQ}=(v_a+s/\sqrt{2})\exp{(i a/\sqrt{2}v_a)}$.

Let us obtain the mass eigenstates in the effective potential of $H_1, H_2, H_0, \AND a$. 
To simplify the discussion we take $A v_a\ll m_H^2$ and $\l v_a^2\ll m_H^2$. 
Then, the mixing between the Higgs fields are suppressed
as follows,
\beq
\theta_{H_{0} H_1}\sim  \frac{A v_a}{m^2_H}\ll 1, \AND~\theta_{H_{0} H_2}\sim  \frac{\l v_a^2}{m^2_H}\ll 1.
\eeq 
{Since $H_2,H_1$ are heavy, the SM-like Higgs is mostly composed of $H_{0}.$}
In this limit, we can neglect the correction to the PQ charge via the mixing of the Higgs fields. The resulting EFT contains the SM fermions with the same quantized PQ charge assignment as in the UV model.
In the unitary gauge, $\vev{H_{0}}\simeq v_{\rm EW}$ is real. Then we cannot use the rotation of the SM Higgs field to remove $a$ in the effective potential, 
\beq
V_{\rm eff}\approx \l v_a^2 e^{i \sqrt{2}a/v_a } H_2 H^*_0+ A v_a e^{i a/\sqrt{2}v_a }H^*_1 H_0+h.c..
\eeq
On the other hand, we can remove  $a$ by redefining $H_1, H_2$. Then,
$a$ appears in the Yukawa terms,  e.g. \beq 
\laq{Yukawaax}y_{u,11} H_2 Q_1 u_1 \to y_{u,11} H_2 e^{-i \sqrt{2}a/v_a} Q_1 u_1,  y_{d,11} H^*_1 Q_1 d_1 \to y_{d,11} H^*_1 e^{-i a/\sqrt{2}v_a} Q_1 d_1 \eeq
and the kinetic terms of $H_{1,2}$. In addition, we can remove $a$ from the Yukawa terms via the chiral rotation of 
$10_{1,2}$ completely. Then we arrive at the form of \Eq{derivative} with the desired charge assignment. 
From the tad-pole conditions, we obtain, 
\beq
\vev{H_1}\simeq  \theta_{H_{0} H_1} v_{\rm EW},~~\vev{H_2}\simeq  \theta_{H_{0} H_2} v_{\rm EW}.
\eeq

The quark flavor structure can be obtained following the MFV assumption.  For instance, we can take the diagonal up-type quark Yukawa matrix, 
$ {Y}_{10;IJ} \theta_{H_0 H_2} = \delta_{IJ} {y}_{uJ}$ for $I,J=u,c$,
and otherwise $ {Y}_{10;IJ} =\delta_{IJ}\, {{y}_{uJ}}$. Then, ${Y}_{5;IJ} \theta_{H_0 H_1} = y_{dI}(V_{\rm CKM}^{-1})_{IJ}$ for ${I,J}=u,c$, and  otherwise ${Y}_{5;IJ}  = y_{dI}(V^{-1}_{\rm CKM})_{IJ}$. {Here, we replaced the indices in the l.h.s to be the flavor one, which is allowed due to the assumption of the alignment;
 $y_{u; IJ}=\delta_{IJ} y_{uJ}$} and $y_{d; IJ}= y_{dI}(V^{-1}_{\rm CKM})_{IJ} $ are the SM Yukawa matrices in the SM Lagrangian,
\beq
{\cal L}\supset - \sum_{I,J=1}^3\(y_{u;IJ}\tilde{H}^* Q_I \bar{u}_J+y_{d;IJ}H^* \bar{d}_I Q_J \) + {h.c.},
\eeq
where $H$ and $\tilde H$ are the SM-Higgs doublet and its conjugate, respectively, {and we explicitly show the bars on $\bar{u}_J$ and $\bar{d}_I$ as in the conventional notation for the Yukawa interaction. }
This model can explain the smallness of the up, down, charm, strange, $\mu, \tau$ masses compared to the top quark mass if $|\theta_{H_0 H_1}|, |\theta_{H_0 H_2}|< 1$. However, the quark-lepton mass relation, especially the small 
electron mass compared to the bottom mass, cannot be explained naturally. Note that, as we have seen in the previous section, the right-handed electron must be in $10_3$.
One solution to these problems is to use the GUT-breaking correction. 
One can introduce higher dimensional operators such as
\beq
\laq{hg}
\delta {\cal L} \supset \frac{\vev{\F_{\rm GUT}}}{M_{*}} H \bar{L}_e e_R + h.c.,
\eeq
where $\F_{\rm GUT}$ is the GUT Higgs field responsible for the GUT breaking,  $M_*$ is the cutoff scale, {and $e_R$ denotes the right-handed electron.}
{This term could cancel the electron Yukawa coupling and reduce it to the observed value, changing the bottom-electron mass relation predicted by the GUT.\footnote{{Below the GUT scale, the electron Yukawa coupling is cancelled to be small and it is stable under the renormalization group running.} 
{On the other hand, above the GUT scale, the cancellation may not be stable under the renormalization group running. We may set the cutoff scale to be very close to the GUT scale or consider the anthropic selection to justify the cancellation of the coupling at the low energy. }} This cancellation might be understood in terms of the anthropic selection~\cite{Agrawal:1997gf}.
}
For $x=1,2$, the renormalization group running between $v_a$ and the GUT scale may change the mass relations for $\{\tau, s\}$, and $\{\m, d\}$ to obtain the factor $\O(10)$ hierarchy if the corresponding $Y_{5}$ is of order unity in the PQ symmetry phase.

In the low-energy EFT obtained by integrating out $\F_{\rm PQ}$ and $ {H_{1,2}}$, we have the desired PQ charge assignment of the fermions. 
{However, the chiral anomaly is different from the hadrophobic conditions discussed in the previous section.} {Indeed, for the above matter content and charge assignment, the anomaly coefficient would be
{\beq
\frac{c_5}{3} \left(2 \times 3\right) = 2 c_5 \ne c_5
\eeq
(or we may also focus on the color subgroup ($\frac{c_3}{3} (2\times 2+1\times 2)=2c_3$).}}
{To match the anomaly coefficient we need additional fields that carry the PQ charge, such as}
\beq
{\cal L}\supset \F_{\rm PQ} \bar{10}_{\rm PQ} 10_{\rm PQ} + {\rm h.c.} 
\eeq
with ${10_{\rm PQ},\bar{10}_{\rm PQ}}$ being a pair of extra PQ fermions {which have the total PQ charge $-c_5
/3$.}
{With the $\F_{\rm PQ}$ VEV, $10_{\rm PQ} \AND \bar{10}_{\rm PQ}$ become massive. Integrating them out induces an anomalous coupling to the gauge bosons, similar to the KSVZ axion. Since this is a complete GUT multiplet, the resulting axion gauge couplings are universal.}

\subsection{A vector-like fermion model}
We provide an alternative UV realization that does not involve the doublet-triplet splitting problem of the additional Higgs fields. We introduce PQ fermions instead of additional bosonic fields to describe the Yukawa interactions. The PQ charge assignment is 
\begin{center}
  \begin{tabular}{@{} cccccccccc @{}}
    \hline
Fields &  $\F_{\rm PQ}$&$\bar{5}_{{\rm PQ};1,2}$&$5_{{\rm PQ};1,2} $&$ \bar{10}_{{\rm PQ};1,2}$&$10_{{\rm PQ};1,2}$&$ 10_{1,2}$&$ 10_{3}$&$ \bar{5}_{1,2,3}$ &$5_H$\\ 
    \hline
    PQ charge$\times 3/c_{5}$  & 1 & -1 & 0 & 0 & -1 & 1 & 0&0 & 0 \\ 
    \hline
  \end{tabular}
\end{center}
where the extra PQ fermions as well as the PQ Higgs are indicated with the subscript ``PQ".

We consider the renormalizable potential for the PQ Higgs given by
\beq\laq{mix2}
{V}= 
 -m_\F^2 |\Phi_{\rm PQ}|^2 + \frac{\l_\F}{2} |\F_{\rm PQ}|^4,
\eeq
which leads to $\vev{\F_{\rm PQ}}=v_a.$ In the EFT after integrating out the PQ Higgs we can replace $\F_{\rm PQ}=v_a \exp(i a/\sqrt{2}v_a).$
The fermion interactions allowed by the symmetry include the following Yukawa couplings,
\begin{align}
\laq{Yukawa3}
{\cal L}_{\rm Yukawa}=&
\sum_{xy=1,2}Y_{5, xy }  \F_{\rm PQ} \bar{5}_{{\rm PQ},x} 5_{{\rm PQ},y}+M_{5,xy}\bar{5}_{x} 5_{{\rm PQ},y} + Y_{d,xy} 5^*_H \bar{5}_{{\rm PQ},x} 10_{y}\\ 
\laq{Yukawa3-1}
+&\sum_{xy=1,2}Y_{10, xy }  \F_{\rm PQ} \bar{10}_{{\rm PQ},x}10_{{\rm PQ},y}+Y_{{\rm 10 mix},xy}\F^*_{\rm PQ}\bar{10}_{{\rm PQ},x}10_{y}  + Y_{u,xy} 5_H 10_{x} 10_{{\rm PQ},y} \\
\laq{Yukawa3-2}
+&\sum_{x=1,2}M_{10,x3}\bar{10}_{{\rm PQ},x} 10_3+M_{5,3x}\bar{5}_3 5_{{\rm PQ},x} +Y_{d,x3}5^*_H\bar{5}_x 10_3\\
+&Y_{d33}5^*_H\bar{5}_3 10_3 +Y_{u33}5_H10_3 10_3.
\end{align}
We have not included interactions such as 
 $5_H 5_{\rm PQ} \bar{10}_{\rm PQ}, 5^*_H \bar{10}_{\rm PQ} \bar{10}_{\rm PQ}$,  which are allowed by symmetry but are not important for our discussion.
The couplings of these irrelevant interactions are assumed to be small enough to simplify the discussion.
We can redefine the fields so that all the phases of $\exp{(i a/\sqrt{2}v_a)}$ are removed and the axion has only  derivative couplings. Then we have the desired form for $10_x $ and $\bar{5}_y$ for the 
{\it model 1}. 
 Assuming $y_5\vev{\F_{\rm PQ}}\gg M_5, y_{10}\vev{\F_{\rm PQ}}\gg M_{10}, y_{10\rm mix} \vev{\F_{\rm PQ}}$, 
 we get a mixing between $\bar{5}_x$ and $\bar{5}_{{\rm PQ},y}$, and  between $10_x$ and $10_{{\rm PQ},y}$, 
 whose mixing parameters are \beq \theta_{5}\sim \frac{M_5}{Y_{5}\vev{\F_{\rm PQ}}} \AND \theta_{10}\sim \frac{Y_{10,{\rm mix}} \vev{\F_{\rm PQ}^*}}{Y_{10}\vev{\F_{\rm PQ}}}, \frac{M_{10}}{Y_{10}\vev{\F_{\rm PQ}}}, \eeq
respectively. Here we have omitted the flavor index. 
 
We can integrate out BSM particles to obtain the SM Yukawa couplings.  This is done by replacing the PQ fermion with mixed SM fermions in the last terms of \Eqs{Yukawa3} and \eq{Yukawa3-1},
\beq
\delta {\cal L}_{\rm eff} \sim \frac{M_5}{Y_5 \vev{\F_{\rm PQ}}} 5_H^* 10_{1,2} \bar{5}_{1,2,3} +\frac{Y_{10\rm mix}\vev{\F^*_{\rm PQ}}}{Y_{10} \vev{\F_{\rm PQ}}} 5_H 10_{1,2} {10}_{1,2}+\frac{M_{10}}{Y_{10} \vev{\F_{\rm PQ}}} 5_H 10_{1,2} {10}_{3},
\eeq
where we have used  short handed notations, and ignored the indices in the Yukawa and mass matrices. 
The first term involves the 3 generations of $\bar{5}$ because, $5_{{\rm PQ},1,2}$ can mix with all of $\bar{5}$ (see the 2nd terms of \Eqs{Yukawa3} and \eq{Yukawa3-2}). Again we can have a generic 3 by 3 Yukawa matrix for the down-type quark $y_{d;IJ}$  and thus the flavor structure can be obtained.   Also the induced Yukawa component by the fermionic mixings are naturally suppressed. 

{For the desired gauge anomalous coefficient for the hadrophobic axion as discussed in the previous section,}
we need additional vector-like fermions, $\bar{5'}_{\rm PQ},5'_{\rm PQ}$, with the coupling of
\beq
\F_{\rm PQ} \bar{5'}_{\rm PQ}5'_{\rm PQ}. 
\eeq
We have checked that the gluon coupling is still asymptotically free with these additional quarks and thus the gauge coupling unification remains perturbative.\\

Before concluding this section, we would like to point out that both models considered above have fine-tuning and hierarchy problems between the EW scale, the PQ scale, and the GUT scale. These issues may suggest a suitable UV completion for the models, such as a SUSY extension. Although we do not present a specific UV completion in this paper, we note that it may be straightforward if the additional scalars/fermions in the two models possess a mass scale relevant to the cutoff scale of the effective theory, for example, the soft SUSY breaking  scale. In this scenario, the axion decay constant is approximately the cutoff scale (see a related model in~\cite{Takahashi:2019qmh}). Interestingly, the resolution of the hierarchy problem between the cutoff scale and the EW scale favors a small axion decay constant, making our discussion of the hadrophobic axion significant.

{More generally, if the decay constant of the light axion originating from the $\SU(5)\times \U(1)_{\rm PQ}$ is related to the cutoff scale to which the EW mass is sensitive, and if the axion couples to the gauge fields, the solution of the hierarchy problem would favor hadrophobic and electrophobic axions. 
}

\section{Axionic unification of inflaton and DM consistent in GUT context} 
\lac{mir}

The idea of unifying the inflaton and DM goes back to the seminal papers~\cite{Kofman:1994rk,Kofman:1997yn}, and has been studied in recent works as well~\cite{Mukaida:2014kpa,Bastero-Gil:2015lga,Chen:2017rpn, Daido:2017wwb, Daido:2017tbr,Hooper:2018buz,Borah:2018rca, Manso:2018cba,Rosa:2018iff,Almeida:2018oid,Choi:2019osi}.
Here we consider the ALP miracle, a unified explanation of inflaton and DM using axion~\cite{Daido:2017wwb, Daido:2017tbr}.

In the ALP miracle scenario, ALP was assumed to interact primarily with photons (and the SM fermions) at low energies and not to couple to gluons. This is because the astrophysical bounds based on the SN 1987A and NSs severely restrict the coupling of ALP to gluons. For this reason, it was considered that the ALP cannot have a universal coupling to SM gauge bosons, and therefore this scenario is difficult to embed in the GUT. Now we know that, as discussed in the previous sections, hadrophobic axion is possible in the GUT framework. Based on this result, we will reconsider the ALP miracle scenario in terms of hadrophobic axion.
In this section, we use $\f$ instead of $a$ to denote the ALP to facilitate comparison with previous literature.

\subsection{The original ALP miracle scenario}

First, let us briefly review the basic idea of the ALP miracle scenario. See Refs.~\cite{Daido:2017wwb, Daido:2017tbr} for details.
In the original model,
the ALP is assumed to have a potential of the form,
\begin{align}
\label{eq:DIV} 
V_{\rm inf}(\phi) = \Lambda^4\(\cos\(\frac{\phi}{f_\f} + \Theta \)- \frac{\kappa }{n^2}\cos\(n\frac{\f}{f_\f }\)\) + {\rm const.}
\end{align}
where $n$ ($>1$) is an integer and $\kappa$ and $\Theta$ represent the relative amplitude and phase of the two terms, respectively. 
We assume that the potential in  \eq{DIV} arises from some UV physics other than QCD, and the parameters in the potential are constant in time.
For successful inflation, $\Theta$ must be much smaller than unity,\footnote{In a certain UV completion based on extra dimensions, $\Theta$ is exactly zero~\cite{Croon:2014dma,Higaki:2015kta,Higaki:2016ydn}.} and $\kappa$ must be close to unity. Then the potential is very flat near the origin, and slow-roll inflation takes place.
In the next subsection, we will introduce the ALP coupling to gluons, which induces another potential term.

The coefficient of $\f/f_\f$ in the first cosine term defines $f_\f$.\footnote{
\label{ft2}We can extend the model so that the first cosine term in \Eq{DIV} contains
another positive integer $n' < n$. 
{By  redefining $f_\phi$, it can be rewritten in the same form as \Eq{DIV}. Then,
 $n$ as well as  $c_{5}$ and $q_{10_1}$ become a rational number, in general.} 
} The dynamical scale $\L$ sets the inflation scale, and it is fixed at ${\cal O}(10)$\,TeV as explained below.
The constant term is added to ensure the vanishingly small cosmological constant in the present vacuum. We introduce two cosine terms in the potential because natural inflation~\cite{Freese:1990rb,Adams:1992bn}, which has a single cosine term, is now excluded due to the null observation of the primordial tensor mode~\cite{BICEP:2021xfz} (see also \cite{Akrami:2018odb}). The next simplest possibility of axion inflation is multi-natural inflation~\cite{Czerny:2014wza, Czerny:2014xja,Czerny:2014qqa,Higaki:2014sja}, which can be realized in certain UV models with extra dimensions~\cite{Croon:2014dma, Higaki:2015kta, Higaki:2016ydn}. Our potential takes the minimal form of the multi-natural inflation.

In this paper, we mainly focus on the case where $n$ is odd, which implies an upside-down symmetry of the potential {under the shift of $\phi \to \phi + \pi f_\phi$:}
\beq
\label{eq:udsym}V_{\rm inf}(\phi)=-V_{\rm inf}(\phi+\pi f_\f )+{\rm const.}
\eeq
As a consequence of this symmetry, the inflaton mass squared at the potential minimum is equal to the curvature at the potential maximum, but with the opposite sign. We do not consider here the case of even $n$, and refer the reader to Refs.\,\cite{Takahashi:2019qmh, Takahashi:2021tff} for discussions of this case. 

This upside-down symmetric potential makes the inflaton so light at the potential minimum that its mass is comparable to that of the Hubble parameter in inflation. Inflation occurs in the flat region near the potential maximum, $\phi \simeq 0$. The inflation dynamics is similar to the hilltop quartic inflation. When the inflation ends and the axion oscillates about the potential minimum, $\phi/f_\phi \simeq \pi$, its couplings to photons and to fermions cause reheating. For the reason explained above, a coupling with gluons is not included in the original set-up. 

The first stage of the reheating occurs due to perturbative decay of the inflaton into light particles such as photons, which quickly thermalize to form a hot plasma. These particles in the interacting plasma give screening or thermal masses to the light particles. In the second stage, the perturbative decay is kinematically forbidden due to the backreaction when the thermal mass becomes comparable to the inflaton. Afterwards, the reheating undergoes with scattering and the inflaton condensate efficiently evaporates {if the evaporation rate is higher than the Hubble parameter.} 
A characteristic feature of this set-up is that the smaller the oscillation amplitude of the inflaton, the lighter the effective mass of the inflaton, and thus the smaller the reaction rate of this reheating process.
For a certain inflationary scale, the rate for the reheating via scattering and the Hubble parameter become comparable at the beginning of reheating. Then,  while most of the energy of the inflaton is used to thermalize the SM particles by reheating (specifically, the evaporation process), some of the inflaton condensates remain intact. This {remnant of the inflaton condensate} becomes DM when the inflaton becomes non-relativistic. The requirement to account for both inflation and DM at the same time determines the inflationary scale of ${\cal O}(10)$\,TeV, and the corresponding Hubble parameter of about ${\cal O}(1)$\,eV.
It is interesting to note that this corresponds to the axion mass of ${\cal O}(0.01-1)$\,eV and an axion-photon coupling $g_{a\gamma \gamma}$ of $10^{-12} - 10^{-10}$\, GeV$^{-1}$, which satisfy the current limits and can be explored by future experiments~\cite{Daido:2017tbr,Daido:2017wwb}.
In such a minimal setup, where inflation occurs with a single axion, it is highly non-trivial that there exists a region that explains inflation and DM simultaneously, and satisfies all the constraints of current observations, and can be explored by future observations. For this reason, this scenario is called the ALP miracle.

\subsection{
ALP potential by QCD
}

The main difference between our model and the original ALP inflation described in the previous subsection is that the ALP has a universal anomalous coefficient with respect to the SM gauge group. In particular, the ALP is coupled to gluons. Thus, there is also the QCD contribution to the potential as given by \Eq{inst}.
We restate the potential due to the non-perturbative effect of QCD once again to match the current notation:
\begin{align}
\laq{inst2}
V_{\rm QCD}(\phi)&= -B_0 f_\pi^2 \sqrt{m_u^2+m_d^2+2 m_d m_u \cos{\left[c_3 \frac{\phi}{f_\phi}\right]}},
\end{align}
where we assume
\beq 
f_\f = \sqrt{2} v_a
\eeq 
to simplify our argument.
The total potential is given by
\beq
V=V_{\rm inf}(\phi)+ V_{\rm QCD}(\phi+(\theta_{\rm QCD} -\pi) f_\f),
\eeq
where we include a relative phase of the two terms, $\theta_{\rm QCD}-\pi f_\f$, which is generically present. With this definition, $\f =\pi f_\f$ is the minimum of $V_{\rm QCD}$ when $\theta_{\rm QCD}=0$. The strong CP phase is given by
\beq
\laq{thetabar}
\bar{\theta}_{\rm CP}\equiv \vev{\phi}/f_\phi-\pi+\theta_{\rm QCD},
\eeq
where $\vev{\phi}$ is the VEV of $\phi$ in the present vacuum.
When $\vev{\f}\approx \pi f_\f$, $\theta_{\rm QCD}$ can be interpreted as the strong CP phase. 
{We will} neglect the finite temperature effect during inflation, since we focus on the case in which the Gibbons-Hawking temperature, $T_{\rm GH}=H_{\rm inf}/2\pi$, is much lower than the QCD scale, $T_{\rm GH}\ll \chi^{1/4}$.

As we have seen in the previous section, in the absence of $V_{\rm QCD}$,  the inflation takes place near the origin, $\phi \approx 0$, and it is stabilized at $\phi/f_\phi \approx \pi$ after inflation. The question is whether the inflaton dynamics in the original setup is modified by $V_{\rm QCD}$. We will see that $V_{\rm QCD}$ does not significantly change the inflaton dynamics. This is essential because its potential height is much smaller than the inflaton potential,
$V_{\rm QCD} \ll V_{\rm inf}$. 
However, as we will see, the curvature of $V_{\rm QCD}$ could change the location of the potential minimum {and maximum}, and thus the strong CP phase.

\begin{figure}[!t]
\begin{center}  
\includegraphics[width=110mm]{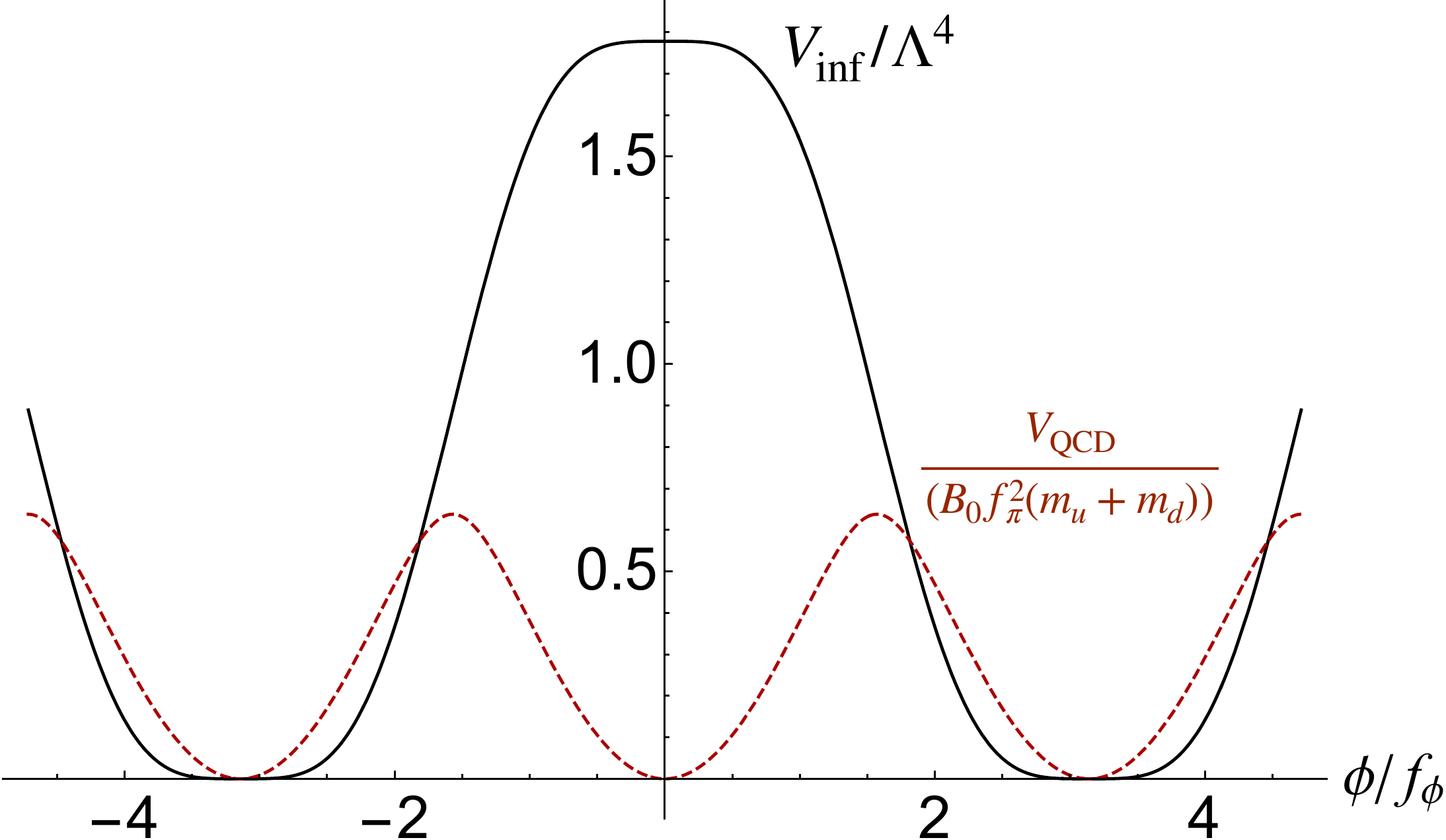}     
      \end{center}
\caption{$V_{\rm inf}/\L^4$ and $V_{\rm QCD}/(B_0 f_\pi^2 (m_u +m_d))$ are shown {as a function of $\f/f_\f$ in black solid and red dashed lines, respectively, 
where we set $n=3$ and $c_3=2$}. For illustrative purposes  we set $\Theta=0$ and $\theta_{\rm QCD}=0$. 
{A constant term is added to adjust the potential at the minima equal to zero.}
}
\label{fig:totalpot} 
\end{figure}

{In the presence of $V_{\rm QCD}$,}
the upside-down symmetry no longer holds, {regardless of the}  parity of $c_3$ (see \Eq{inst2}). In Fig.~\ref{fig:QCD} we plot  $V'_{\rm QCD}(\f)+V'_{\rm QCD}(\f +\pi f_\f)$ (left panel) and $V''_{\rm QCD}(\f)+V''_{\rm QCD}(\f +\pi f_\f) $ (right panel)  in black solid and red dashed lines, for $c_3=1$ and $c_3=2$, respectively. The deviation from zero of these functions implies that the upside-down symmetry is explicitly broken by $V_{\rm QCD}$.
{As a consequence, the first derivative of the potential flips the sign under the shift of $\phi \to \phi + \pi f_\phi$ only at}
{$\f/f_\f= 0 ~{\rm mod}~\pi/2$}. This property will be important for a consistent DM cosmology.

\begin{figure}[!t]
\begin{center}  
\includegraphics[width=80mm]{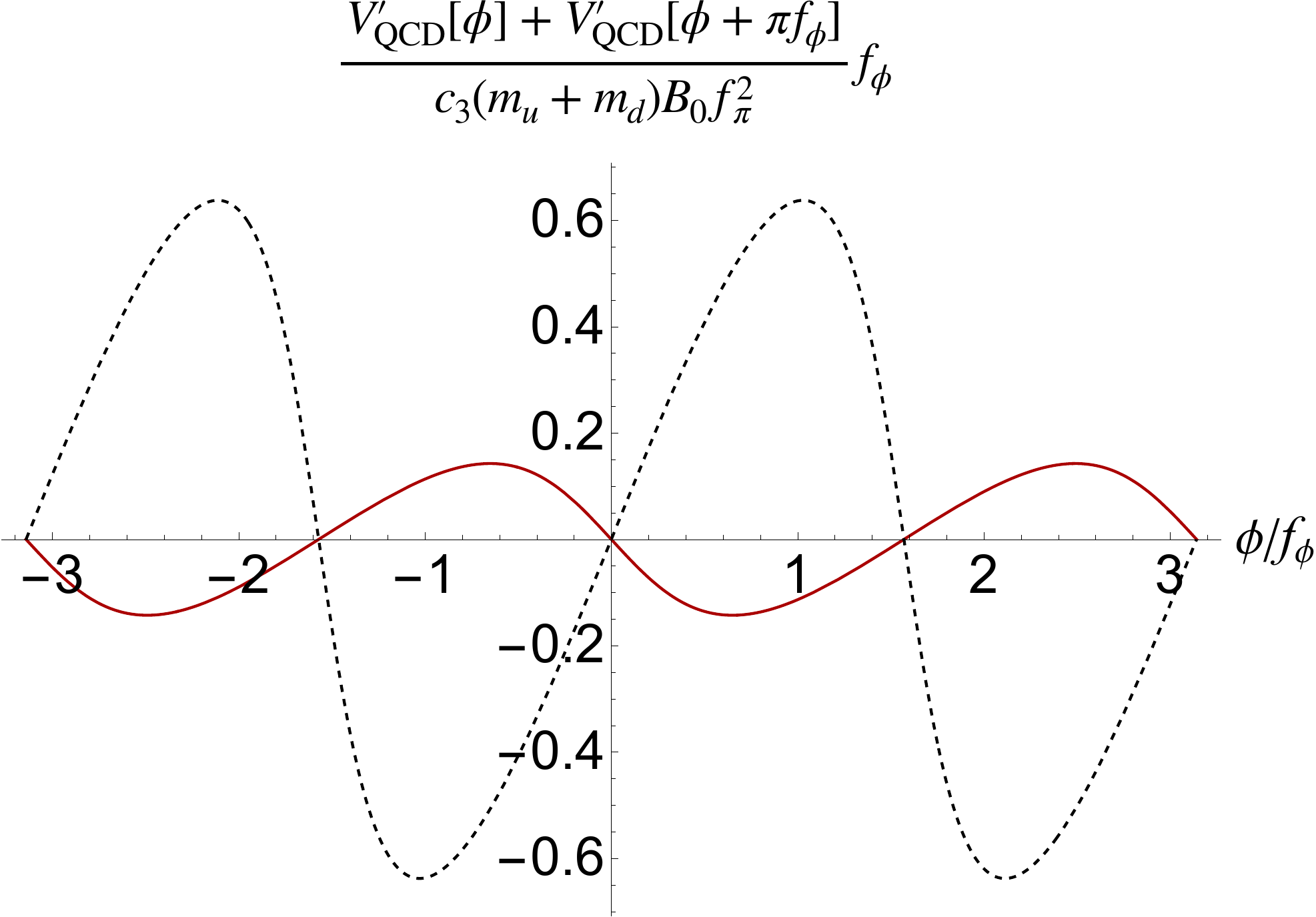}     
\includegraphics[width=80mm]{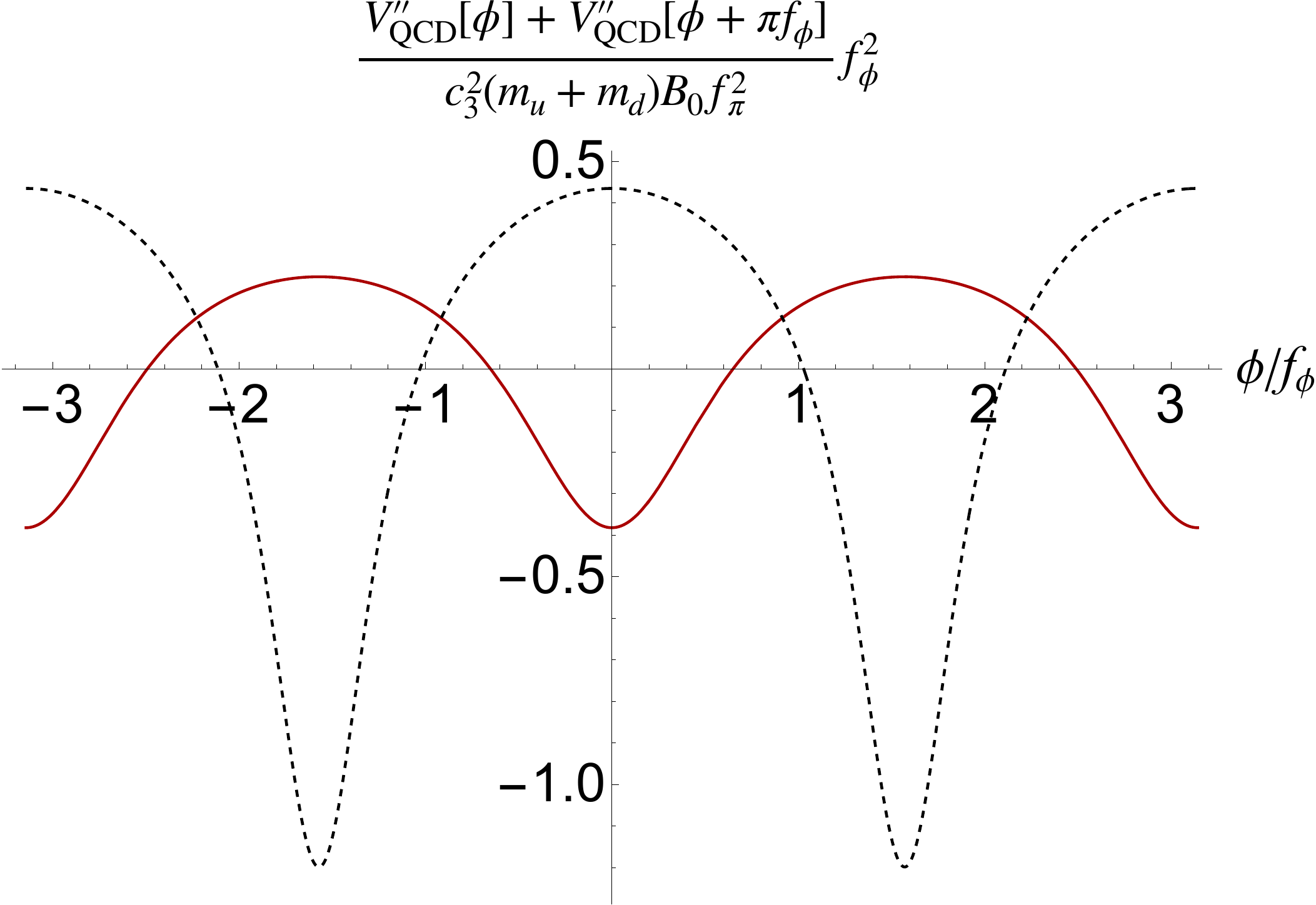}     
      \end{center}
\caption{
The breaking of the upside-down symmetry induced by the QCD instanton is shown in the red solid and black dashed lines for $c_3=1$ (odd) and $2$ (even), respectively.
In the left and right, we plot  $ (V'_{\rm QCD}(\f)+V'_{\rm QCD}(\f +\pi f_\f))f_\f/(c_3 \chi)$ and $(V''_{\rm QCD}(\f)+V''_{\rm QCD}(\f +\pi f_\f))/(c_3^2 \chi) $,  respectively. 
}\label{fig:QCD} 
\end{figure}

\subsection{Viable parameter region for successful inflation}

For successful inflation, we assume that the inflaton initially stays in the vicinity of the potential maximum near the origin.
Around the origin, the potential can be expanded as
\begin{align}
\label{eq:app}
V(\phi) = V_0   - \h \frac{\L^4 }{ f_\f} {\phi} +\frac{m^2 }{2}\f^2 +A \f^3- \lambda \phi^4 + \cdots    
\end{align} 
with
\begin{align}
V_0 &\simeq \left(2 - \frac{2}{n^2} \sin^2{\frac{n \pi}{2}} \right) \Lambda^4\\
\theta&\simeq \Theta - (\theta_{\rm QCD}-\pi) \frac{\chi}{\L^4} c_3\\
m^2&\simeq \frac{(\kappa-1)}{2}\frac{\L^4}{f_\f^2}-\frac{1}{2}\frac{\chi}{f_\f^2} c_3^2\\
A&\simeq \frac{\H }{3!}\frac{\Lambda^4 }{f_\f^3} - \frac{\theta_{\rm QCD}-\pi }{3!}\frac{\chi }{f_\f^3} c_3^3  \\
\l&\simeq \frac{n^2-1 }{4!}\(\frac{\Lambda}{f_\f}\)^4 ,
\end{align}
where we have neglected subleading terms. {Note that $m^2 < 0$ in the parameters of our interest. }
For the hilltop inflation, the cubic term has a negligible effect on the inflaton dynamics, and  we have~\cite{Takahashi:2019qmh}
\beq
\label{range}
|\theta| \lesssim \(\frac{f_\f}{M_{\rm pl}}\)^{3},~~ |m^2| \lesssim \frac{\L^4}{f_\f^2}\(\frac{f_\f}{M_{\rm pl}}\)^{2}.
\eeq
for $n=O(1)$.
On the other hand, we need
\beq
\theta = \O\(0.001-0.1\)\(\frac{f_\f}{M_{\rm pl}}\)^3
\eeq
to  explain the observed scalar spectral index of the density perturbation~\cite{Takahashi:2013cxa}.\footnote{
In some concrete UV models with $\Theta = 0$~\cite{Croon:2014dma, Higaki:2015kta, Higaki:2016ydn}, we need $|\theta_{\rm QCD}|\simeq \O(10^{-11}\text{-}10^{-9})\, c_3\(\frac{f_\f}{5\times 10^7{\rm GeV}}\)^8 (\frac{\l}{10^{-12}})$ 
to explain the observed spectral index. {The induced $\bar{\theta}_{\rm CP}$ can be also estimated to be a similar order.} 
This scenario may be tested in future nucleon EDM measurements~\cite{Anastassopoulos:2015ura,Omarov:2020kws}.  On the other hand, $\Theta$ is generically nonzero and contributes to the strong CP phase when the inflaton potential is generated by a small instanton or mirror instanton, in which case a similar prediction can be obtained (but with a different scaling of $f_\f$)~\cite{Takahashi:2021tff}.}
 {Since the quadratic term is small from the slow-roll condition}, by comparing the linear and quartic terms, the location of the potential maximum can be determined as
\beq
\laq{phimax}
\f_{\rm max}= \O\(\frac{f^2_\f}{M_{\rm pl}}\). 
\eeq
This condition does not change much by varying $m^2$ {within the slow-roll regime}. 

 The difference in the inflation potential from the original ALP inflation potential comes from the cubic term {while the additional contributions to the linear and quadratic terms can be absorbed by $\Theta$ and $\kappa$ as can be seen from \Eq{app}.} To assess the contribution of the cubic term to the inflaton dynamics, we recall that one of the slow-roll conditions at the hilltop, i.e. the condition for hilltop inflation, is given by
\beq
H_{\rm inf}^2 \simeq \frac{\L^4}{3 M^2_{\rm pl}} \gtrsim  |V''(\f_{\rm max})|
\eeq
Indeed, the conditions of Eq.\,(\ref{range}) can be obtained by requiring each contribution from the quadratic and quartic terms at the hilltop to satisfy the inequality. We can also restrict {the cubic coupilng} $A$ using this condition. It turns out that
\beq
\frac{|A|}{\L^4/f^2_\f M_{\rm pl}} < \O(1).
\eeq
The QCD potential contributes to the cubic term as 
\beq
\frac{\d A}{\L^4/f_\f^2 M_{\rm pl}}= -10^{-11}(\theta_{\rm QCD}-\pi) c_3^{3}\(\frac{\L}{10^4\GEV}\)^{-4} \(\frac{f_\f}{10^8\GEV}\)^{-1}. 
\eeq 
Since $|\theta_{\rm QCD} -\pi|\lesssim {\cal O}(1)$, this contribution is always negligible in inflationary dynamics. 

In summary, the inclusion of $V_{\rm QCD}$ does not change the inflationary dynamics from the original ALP miracle scenario.
Therefore, we can safely use the {results of the} previous studies~\cite{Czerny:2014wza, Czerny:2014xja,Czerny:2014qqa,Higaki:2014sja, Daido:2017wwb,Daido:2017tbr,Takahashi:2019qmh}, and in particular, the CMB normalization fixes the quartic coupling as
\beq
\l \simeq  10^{-12} ~~~ \text{[CMB normalization]}
\eeq
in the parameter range of interest. {This is a general property of the hilltop quartic inflation.}
Similarly, the observed spectral index can be explained if 
\beq
\label{ns}
|V''(\f_{\rm max})| =\zeta H_{\rm inf}^2~~~~\text{[spectral index]}
\eeq
with $\zeta =\O(0.01-1)$. 
We also note that in the low scale inflation {like the one considered here}, the tensor-to-scalar ratio is predicted to be extremely small, $r \simeq 2\times 10^{-47}\(\frac{H_{\rm inf}}{1\EV}\)^{2}$, which is of course consistent with the current limit, $r<0.036 (95\%{\rm CL})$ ~\cite{BICEP:2021xfz}.

\subsection{ALP mass and the strong CP problem}

Now let us discuss the ALP around the potential minimum. In this case, the QCD instanton contribution becomes important since it breaks the upside-down symmetry. 

The potential around $\phi = \pi f_\f$ is given by 
\beq
V(\phi) \simeq  \tl{\h} \frac{\L^4 }{ f_\f} {(\phi-\pi f_\f)} -\frac{\tl{m}^2 }{2}(\phi-\pi f_\f)^2 -\tl A (\phi-\pi f_\f)^3+ \lambda (\phi-\pi f_\f)^4.
\eeq
If the QCD contribution were neglected, we would obtain $\tl \theta=\theta$, $\tl m^2=m^2$, and $\tl A=A$ because of the upside-down symmetry.
{Note that $\tilde{m}^2 < 0$ in this definition.}
Then the ALP mass at the potential minimum, {$m_\phi$ would be given by} 
\beq
m_\f^2 =  \zeta H^2_{\rm inf}~~~~~[\text{upside-down symmetric limit}],
\eeq
which is fixed by the spectral index (see Eq.~(\ref{ns})).
{Because of this property, the ALP remains light in the present vacuum, and it can be stable enough to become DM.}

However, as we have seen above, the upside-down symmetry 
is explicitly broken by the QCD contribution. 
The linear term {happens to be equal in magnitude} and opposite in sign {at} $\theta_{\rm QCD}=0~~ ({\rm mod}~~ \pi/2)$. See Fig.~\ref{fig:QCD}. In the regions around $\theta_{\rm QCD}\approx \pi/2, \pi, 3\pi/2$, however, the expansion in the above does not work or/and the strong CP phase will be too large to be consistent with the experimental bound. So we will focus on 
the case of $|\theta_{\rm QCD}| \ll 1$.
In this case,
we obtain 
\beq 
\label{thth}
\tl\theta-\theta \simeq \O(0.1-1) \theta_{\rm QCD} \frac{\chi}{\L^4} c_3,
\eeq
{which is induced by a small relative phase between $V_{\rm inf}$ and $V_{\rm QCD}$.}

{In the following we give an analytical estimate of the additional contributions to the ALP mass at the potential minimum. First, the change of the linear term (\ref{thth}) induces an extra contribution to the ALP mass, which comes from the balance between the linear term and the quartic term. Denoting this contribution by $\d_{\rm CPV} m_\f^2$, we obtain}
\beq
\laq{ALPmassCP}
\d_{\rm CPV} m_\f^2 \sim 
 10^{-11}\GEV^2 |\theta_{\rm QCD} |^{2/3} \(\frac{10^{7}\GEV}{f_\f/c_3}\)^{2/3} \(\frac{\l}{10^{-12}}\)^{1/3}.
\eeq
{This contribution is less than $1\KEV$ for $|\theta_{\rm QCD}| \lesssim 1$}
when $f_\f /c_3 \gtrsim 10^7\GEV$. 
Note that this contribution vanishes when $\theta_{\rm QCD} \to 0$.

{Second, the QCD contribution $V_{\rm QCD}$ contributes to the mass,
which does not vanish even when $\theta_{\rm QCD}=0$. In fact, the quadratic terms at $\phi = 0$ and $\phi = \pi f_\phi$ are related as}
{
\beq 
\tl{m}^2=m^2- V_{\rm QCD}''(0)-V_{\rm QCD}''(\pi f_\f).
\eeq}
{The upside-down symmetry breaking contribution to the} quadratic term, $V_{\rm QCD}''(0)+V_{\rm QCD}''(\pi f_\f)= \O(0.1-1) c_3^2 \chi/f_\f^2$, is positive (negative) if $c_3$ is even (odd) when $\theta_{\rm QCD} \to 0$. See Fig.~\ref{fig:QCD}. 
{This is the additional correction to the ALP mass, which arises from the explicit breaking of the upside-down symmetry of $V_{\rm QCD}$. Note that the sign of this contribution depends on the parity of $c_3$. }
When $c_3$ is odd, there might be some cancellation. In particular,  
the ALP mass may be smaller than $c_3^2 \chi/f_\f^2$ by a
{factor of $\xi_T$.}
{This is } due to the cancellation between $\(V_{\rm QCD}''[0]+V_{\rm QCD}''[\pi f_\f]\)$ and other contributions. 

To sum up, the ALP mass is obtained by the largest of the three contributions, 
\beq
\laq{ALPmass}
m_\f =\sqrt{\max[ \z H^2_{\rm inf},  \d_{\rm CPV} m_\f^2, \x_T c_3^2 \chi/f^2_\f ]}.
\eeq
{We set $\xi_T=0.1 (0.43)$ for the odd (even) values of $c_3$. The choice of $\xi_T$ in the even case is based on numerical calculations. In the case of odd values, we have taken into account the cancellation of positive and negative contributions, which amounts to about $10\%$.}\footnote{We have numerically checked that the cancellation cannot be {more significant}. }

In Fig.\ref{fig:CP} we plot the prediction of $\theta_{\rm QCD}$ with respect to $m_\phi$ by stabilizing the potential, 
for $c_3=1$ (red circle points) and $c_3=2$ (black triangle points) in the left and right panels, respectively.
Here we randomly take  $\log_{10}{\theta \(f_\f/M_{\rm pl}\)^{-3}}=[-3,0]$, $\log_{10}[-V''/H_{\rm inf}^2]|_{\rm hilltop}=[-3,0]$,  and $\log_{\rm 10}|\theta_{\rm QCD}|=[-15,-5]$ with both signs. We fix $\l=10^{-12}$ and $n=3$, $f_\phi=10^8\,$GeV. 
 In the lower mass region the contribution of $H_{\rm inf}^2\x$ in \Eq{ALPmass} is dominant, while $\d_{\rm CPV} m_{\f }^2$ is dominant at larger mass scale.
In Fig.~\ref{fig:CP-1} we show the cases for $f_\phi=10^7\GEV, 10^6\GEV$. In this case, the dominant contribution at smaller $|\theta_{\rm QCD}|$ is from $\x_T c_3^2 \chi /f_\f^2$ in \Eq{ALPmass}. 
This contribution is the same as the ordinary QCD axion. It increases when $f_\f$ decreases. For odd $c_3$, there is a mild cancellation between the contributions of $\d_{\rm CPV}m_\f^2$ and $\x_T c_3^2 \chi/f_\f^2$. The contribution of $\zeta H_{\rm inf}^2$  is {negligibly small} in the whole range. In Fig.~\ref{fig:CP-2}, we take $f_\phi=2\times 10^7\GEV$, {in which case there is a } region where all three contributions are important. {We have confirmed that the analytical estimate of the ALP mass given by \Eq{ALPmass}} is consistent with the numerical result.

\begin{figure}[!t]
\begin{center}  
\includegraphics[width=75mm]{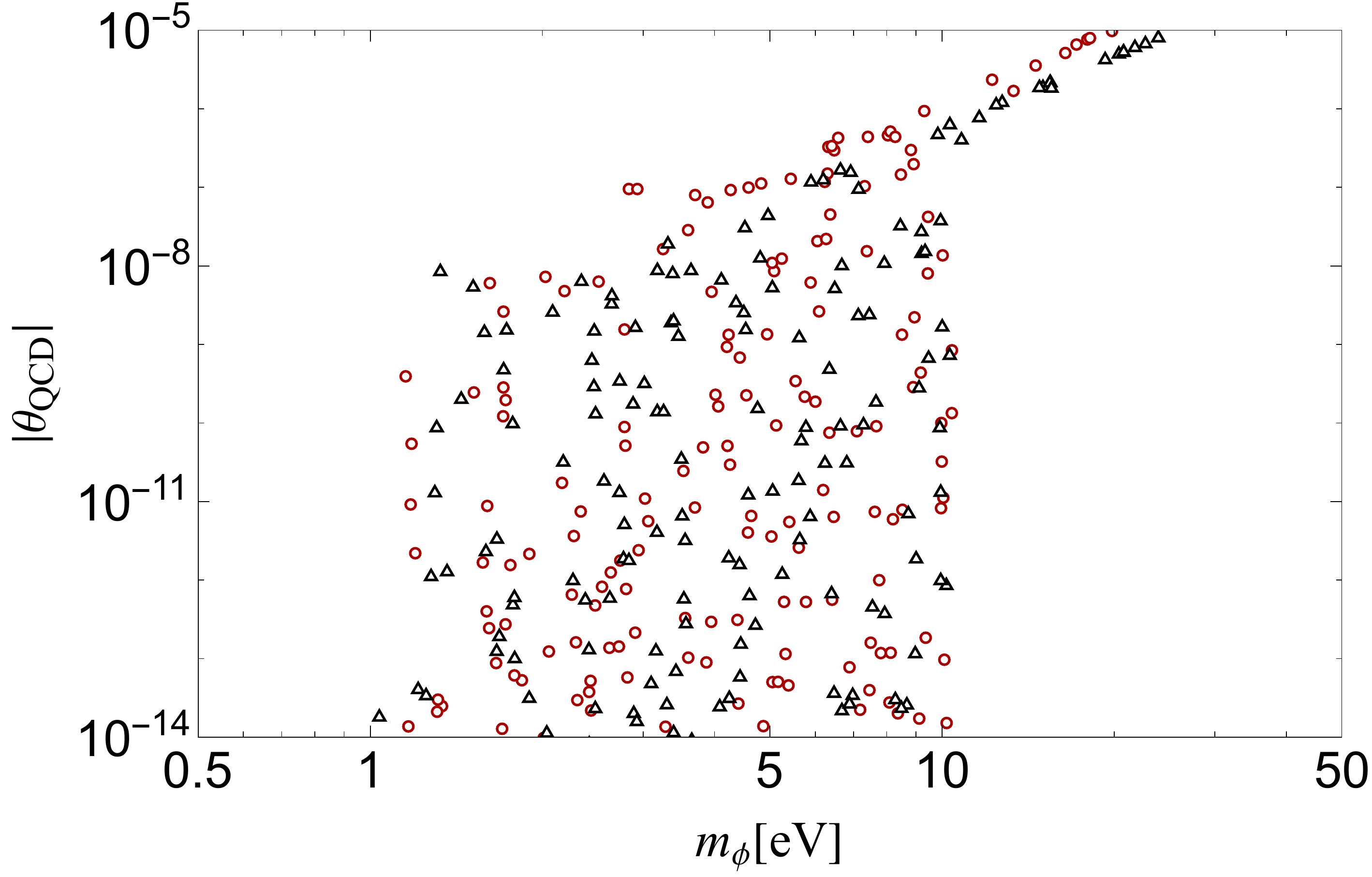}     
\includegraphics[width=75mm]{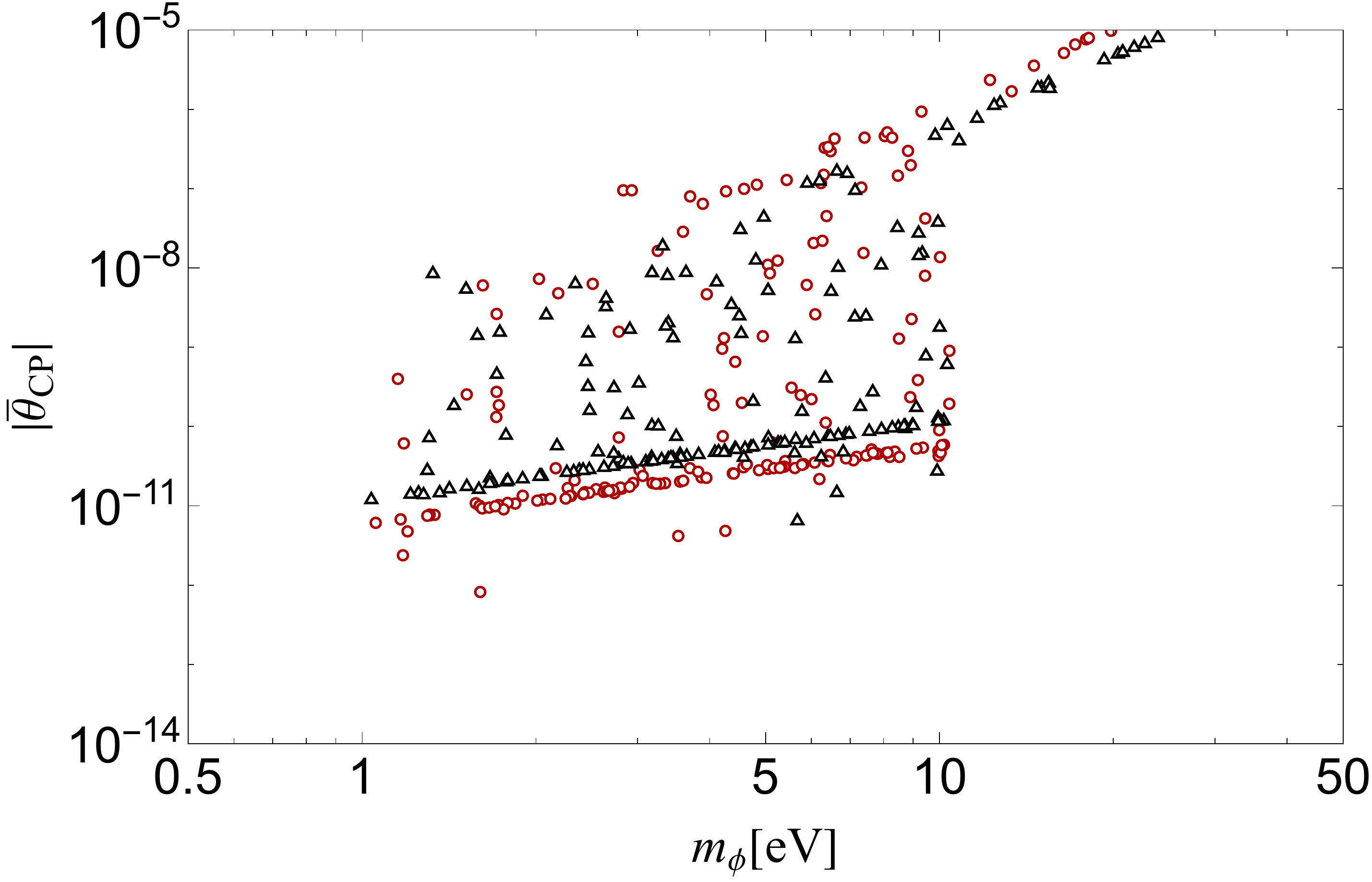}     
      \end{center}
\caption{
{
The relative phase, $\theta_{\rm QCD}$ (left panel) and the strong CP phase, $\bar\theta_{\rm CP}$ (right panel) with respect to the ALP DM mass, $m_\f$, evaluated at the potential minimum. We consider $f_\phi=10^8 \GEV$ with $c_3=1$ (red circles) and $c_3=2$ (black triangles). The hot DM mass limit, $m_\f\lesssim 1 \EV$, constrains the strong CP phase.
}
}\label{fig:CP} 
\end{figure}

\begin{figure}[!t]
\begin{center}  
\includegraphics[width=75mm]{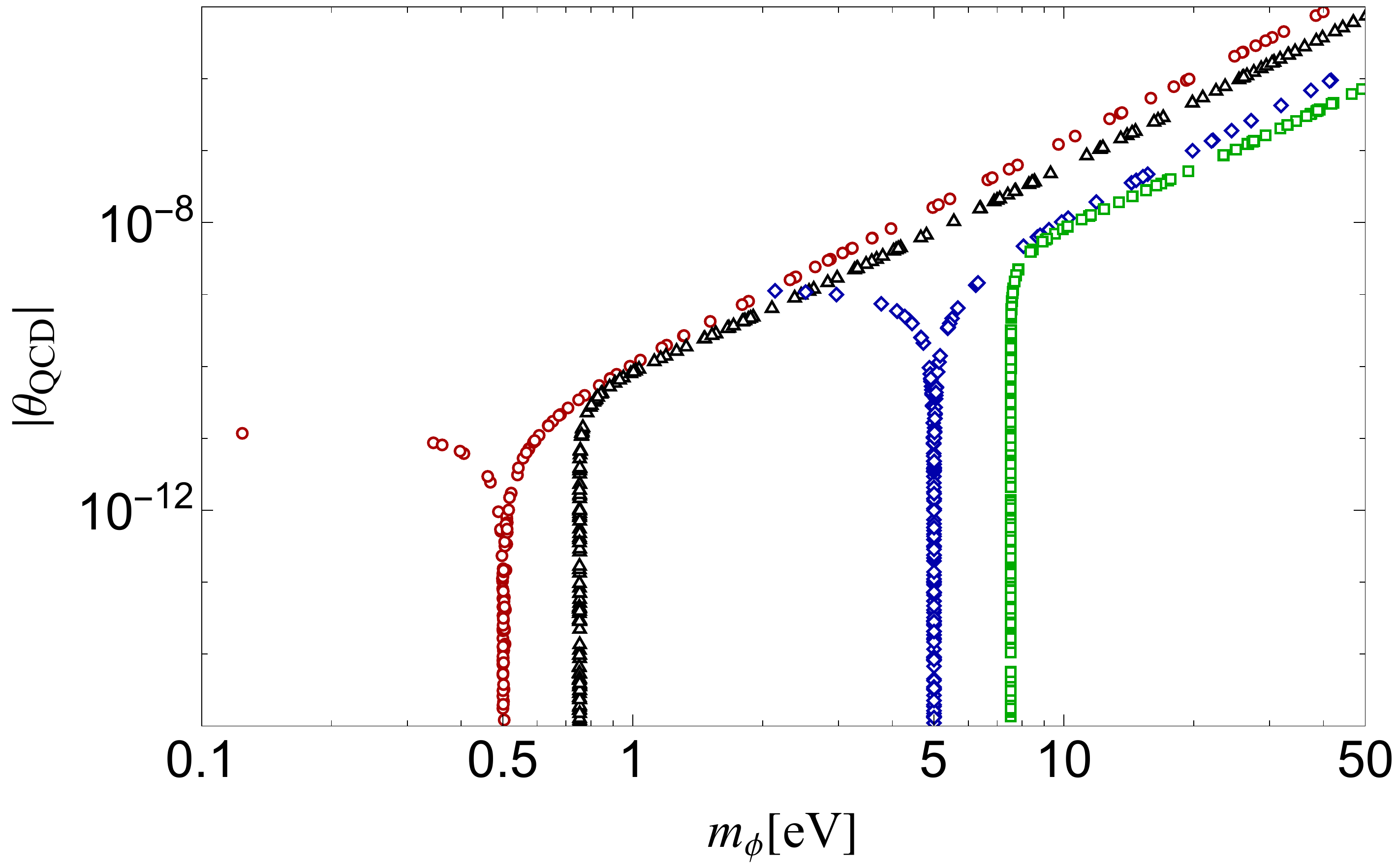}     
\includegraphics[width=75mm]{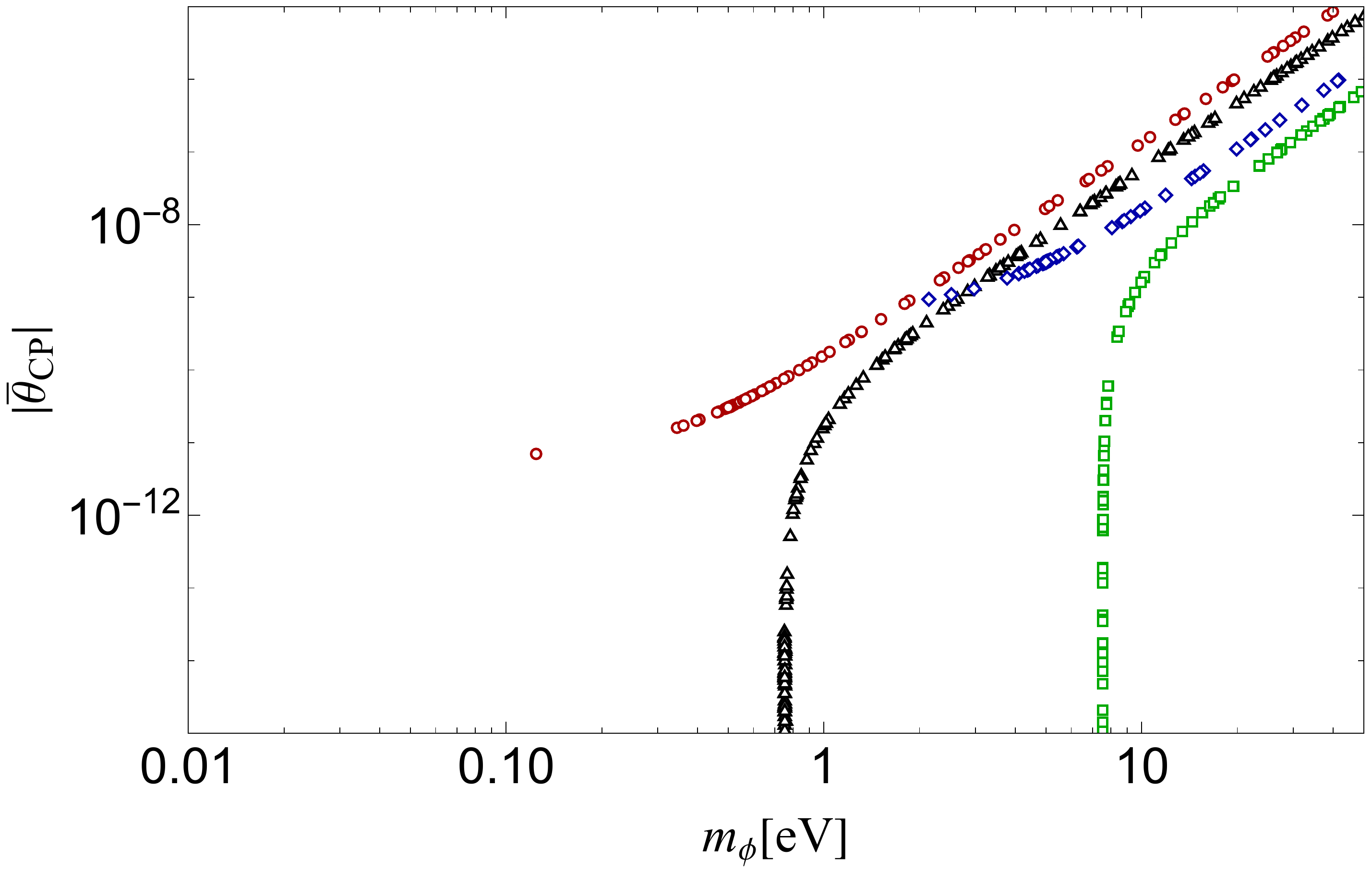}     
      \end{center}
\caption{Same as Fig.~\ref{fig:CP} but with different $f_\f$. Data points of red circles, black triangles, blue diamonds, and green squares represent $(f_\phi, c_3)$ values of $(10^7\GEV, 1), (10^7\GEV, 2), (10^6\GEV, 1)$, and $(10^6\GEV, 2)$, respectively. The hot DM mass limit, $m_\f\lesssim 1 \EV$, constrains the strong CP phase. 
}\label{fig:CP-1} 
\end{figure}

\begin{figure}[!t]
\begin{center}  
\includegraphics[width=75mm]{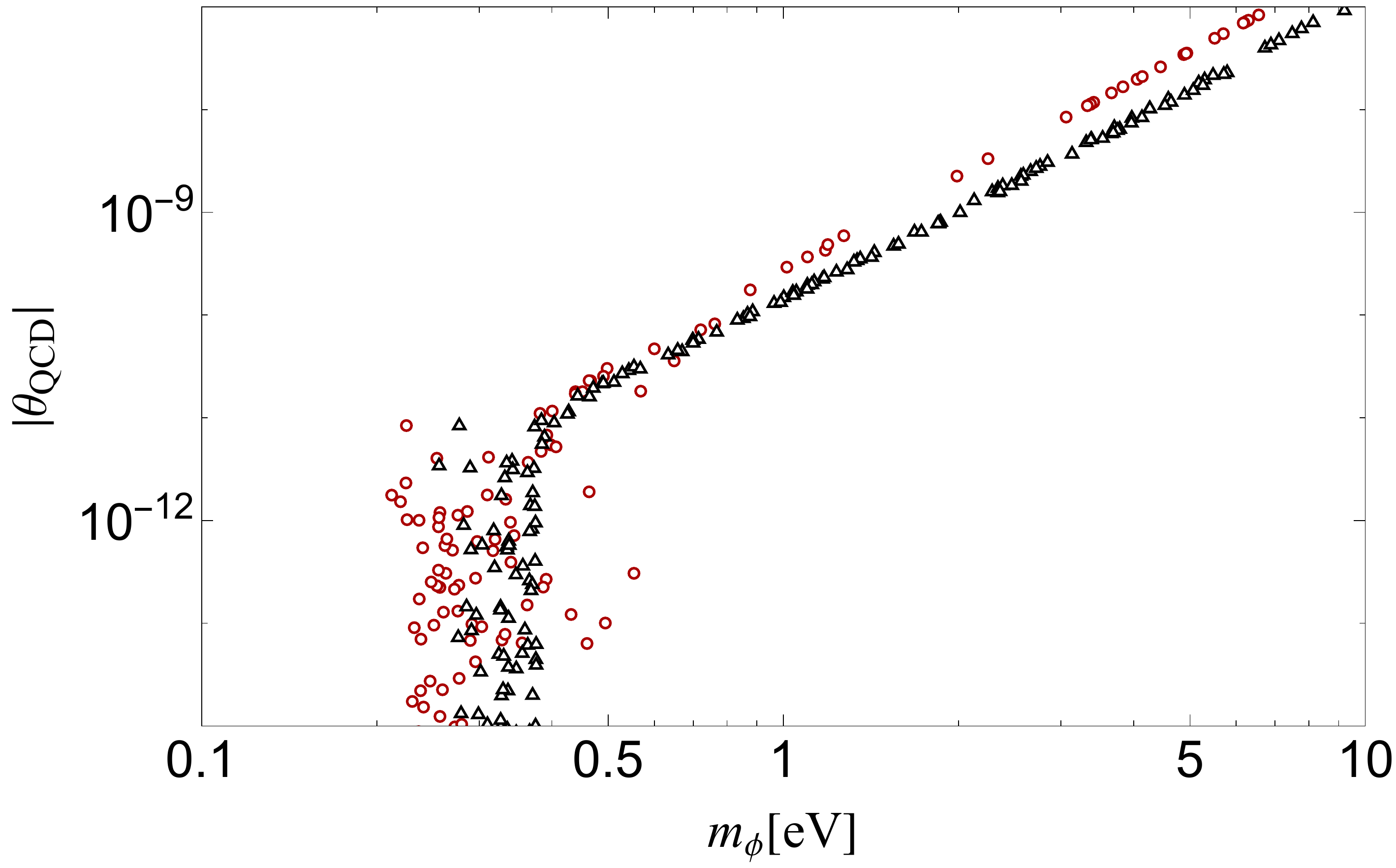}     
\includegraphics[width=75mm]{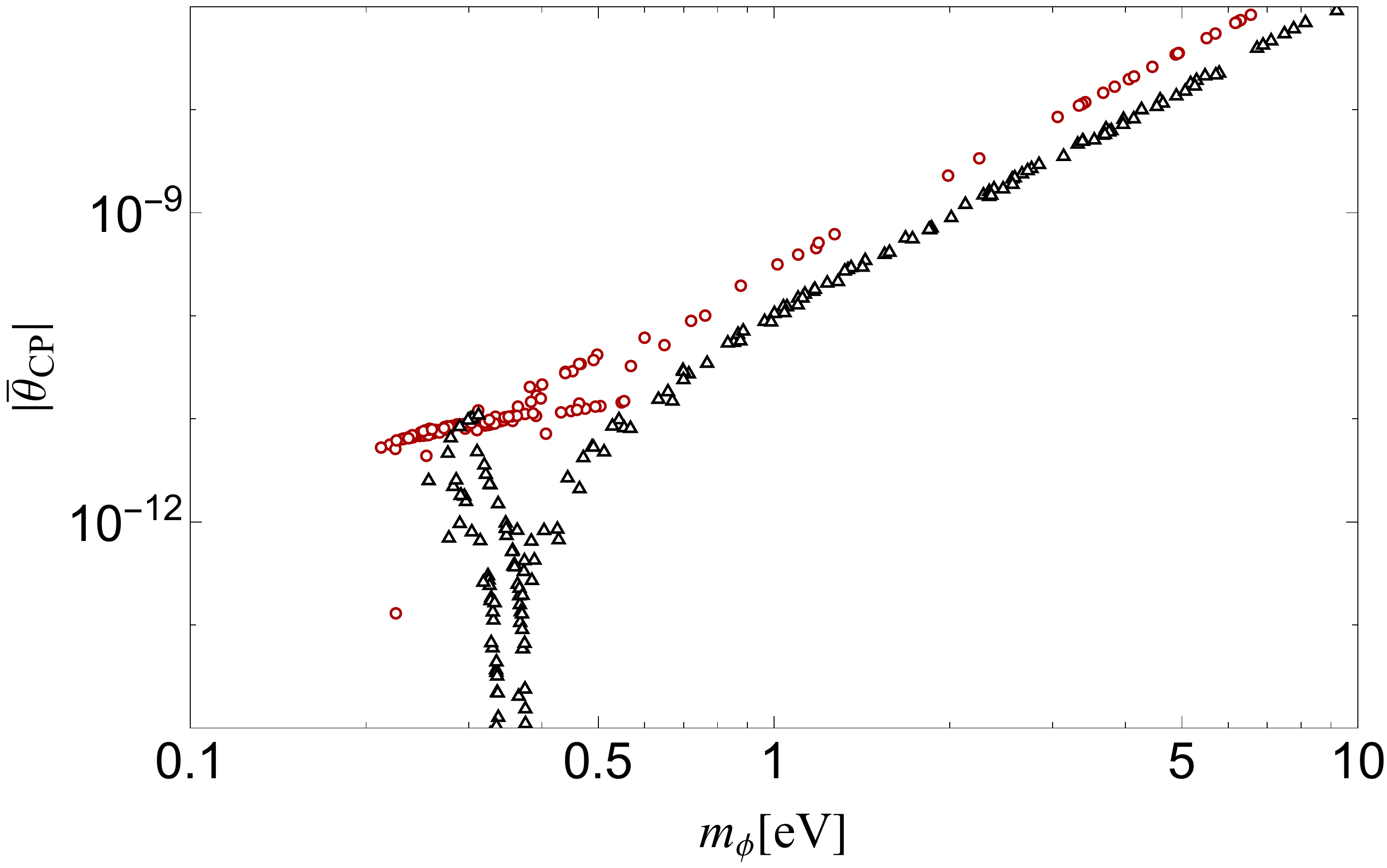}     
      \end{center}
\caption{Same as Fig.~\ref{fig:CP} but with $f_\f=2\times 10^7\GEV$. 
The hot DM mass limit, $m_\f\lesssim 1 \EV$, constrains the strong CP phase. 
}\label{fig:CP-2} 
\end{figure}

As we will see in the next section, during reheating, ALP particles are produced and thermalized, while the ALP condensate remains as cold DM. The ALP particles decouple after reheating and contribute to the dark radiation with
\beq
\D N_{\rm eff}= \O(0.01).
\eeq
The abundance fraction of hot DM places an upper bound on the ALP mass of~\cite{Daido:2017tbr} (translated from \cite{Osato:2016ixc})
\beq
m_{\f}< 7.7 \,\EV \(\frac{0.03 }{\D N_{\rm eff} }\)^{3/4}.
\eeq
Note that the ALP production from ALP-hadron interaction, including the axion-pion mixing, can be neglected due to the hadrophobia compared to the {previously discussed contributions} (c.f.~\cite{Chang:1993gm, Moroi:1998qs}). 
This hot DM mass bound leads to a range for the ALP decay constant, 
\beq
\laq{fphi}
10^6\GEV \lesssim f_\f \lesssim 10^{8}\GEV.
\eeq
The upper bound is from the first term in \Eq{ALPmass}, while the lower bound is from the last term of \Eq{ALPmass}. As we will see, this range is also supported by the requirement for successful reheating and the correct DM abundance.

{
For clarity, the range for $f_\phi$ specified in Eq. \eq{fphi} suggests that the value of $(\langle \phi\rangle/f_\phi -\pi)$ is approximately $\mathcal{O}(f_\phi/M_{\rm pl}) \ll 10^{-10}$. This aligns with the experimental upper limit on the strong CP phase, and is derived from the upside-down symmetry and Eq. \eq{phimax}.
}
The same hot DM bound also {constrains} $\theta_{\rm QCD}$.
{By requiring the mass contribution from \Eqs{ALPmassCP} is less than $1\EV$, we derive}
\beq 
\laq{strongCP}
\boxed{|\theta_{\rm QCD}|,~|\bar{\theta}_{\rm CP}|\lesssim 10^{-9}.}\eeq 
{
This result is intriguing as it suggests the suppression of the nucleon EDM. However, it is crucial to note that this does not imply the absence of any fine-tuning for solving the strong CP problem. Instead, it integrates the fine-tuning into the DM sector. The presence of cold DM in the correct abundance, along with the absence of a hot DM component, is crucial for successful structure formation. Thus, the strong CP problem is anthropically resolved (for similar approaches, see Refs.~\cite{Takahashi:2008pu,Kaloper:2017fsa,Dine:2018glh,Takahashi:2021tff,TitoDAgnolo:2021nhd}).
}

{We find that the lower limit of the ALP mass, as derived from \Eq{ALPmass}, is about $0.01\EV \lesssim m_\phi$. This results from the first (last) contribution being an increasing (decreasing) function of $f_\f$. We note that, if this bound is satisfied,  the structure formation bound for the late-forming  DM is also satisfied~\cite{Daido:2017wwb,Daido:2017tbr}. Thus, the predicted mass range is \beq\laq{massrel} 0.01\EV \lesssim m_\phi\lesssim 1\EV,\eeq
which is also confirmed by our numerical results.}

{So far we have assumed the completion of the reheating via the dissipation effect, which leads to the production of thermal ALPs. Thus, the resulting \Eqs{strongCP} and \Eq{fphi} do not depend on the specific details of the ALP interaction. These conditions help us to identify potential ALP-fermion interactions where the parameter range for successful reheating and DM abundance is consistent with \Eq{fphi}.
}

\subsection{Successful reheating and viable parameter region}
Now let us study the reheating process in more detail.
To satisfy the bound on $f_\phi$ given by \Eq{fphi}, we should focus on the models in \Sec{flavor}.
It is known that the ALP condensate will completely evaporate {and does not explain DM}, when $c_t=\O(0.1-1)$ in the range of $f_\phi$ given by \eq{fphi} \cite{Daido:2017tbr, Takahashi:2020uio}. This leads us to consider the normal embedding of the {\it model 1}, which has $c_t=0$.

By redefining $10_{1,2}$ to have the ALP in the SM Yukawa coupling phases, we obtain\footnote{When we integrate out the 2nd generation quarks, we get the vanishing ALP-gluon coupling and \eq{lag}. }
\beq
{\cal L}_{\rm int}= (2c_{5}-c_5)\frac{\f }{f_\f} \sum_{i=1}^3\frac{g_i^2}{32\pi^2}   F_i\tl{F}_i = c_{5}\frac{\f }{f_\f} \sum_{i=1}^3\frac{g_i^2}{32\pi^2}   F_i\tl{F}_i
\eeq
Then the interacting Lagrangian in the symmetric phase is 
\begin{align}
\laq{lagreh}
{\cal L}_{\rm int}\approx -\frac{i  c_{5}\f }{3f_\f} \times  \(2\sum_{x=1,2} y_{ux}H Q_{x}  {u}_{Rx}
 +y_{dx}H^* Q_{x}   {d}_{R,x}
+y_{e x} H^* L_{x}   {e}_{R,x}\) +h.c.
\end{align}
where we have dropppted terms of  $\O(\f^2)$, and $x$ denotes the GUT index  related to the mass flavor (see the discussion below \Eq{rel}).
In the following we neglect the gauge interactions, whose perturbative dissipation rates are known to be less efficient than the dissipation rates from the ALP interactions with tau and charm~\cite{Daido:2017tbr}.

After the end of inflation, the ALP field starts to oscillate around its minimum with a large effective mass given by
\beq
m_{\rm eff}^2\sim 12\l \f_{\rm amp}^2 = \(35\GEV\)^2 \frac{\l}{10^{-12}} \(\frac{\f_{\rm amp} }{10^{7}\GEV}\)^2.
\eeq
The first step in the reheating process is the perturbative decay of the ALP condensate, $\f\to \bar\psi_L \psi_R$, with $\psi$ being a fermion in the Lagrangian \eq{lagreh}. The decay rate is given by
\beq
\G_{{\rm dec},\psi}\simeq
\frac{N_c}{8\pi } \(\frac{c_\psi m_\psi }{f_\f}\)^2m_{\rm eff}.
\eeq
where $N_c$ is the color factor when $\psi$ is a quark (if $\psi$ is a lepton, $N_c=1$).

The {produced} fermions soon thermalize via the SM $2\to 2$ or $2\leftrightarrow 3$ interactions and form a thermal plasma with temperature $T$.
This temperature gradually increases as the $\f$ decay proceeds.
Within a Hubble time, 
$e T \sim m_{\rm eff}$ is satisfied, and then the back reaction becomes important.

The first back-reaction is the thermal blocking effect, which occurs when the phase space for the decay closes due to the sizable thermal mass, e.g. $m_{\rm lepton}^{\rm th}\sim e T. $
Another effect is the dissipation effect due to the scattering of $\f \psi_L \to \psi_R \gamma$ and other processes. The dissipation rate is given by \cite{Mukaida:2012bz}
\beq
 \G_{{\rm dis},\p}\sim  0.5 N_c \(\frac{c_\psi m_\psi  }{f_\f}\)^2 \(\frac{ \a_\psi }{2\pi^2}\) T.
\eeq
Here, $\a_\psi=\a\approx 1/137$ for leptons and $\a_\psi=\a_s\approx 0.1$ for quarks.
The temperature continues to increase due to this process.\footnote{
See also Refs.\,\cite{Kofman:1997yn,Berera:1995ie,Berera:1998gx,Yokoyama:1998ju,Nakayama:2021avl} for deriving the dissipation term from the equation of motion of the ALP oscillation.}
When the temperature becomes higher than the weak scale, the scattering
$ \f H \to \psi_L \psi_R $ starts to occur, where $H$ is the Higgs boson.
The dissipation rate is given by \cite{Salvio:2013iaa, Daido:2017tbr}
\beq
\laq{disH}
\G_{\rm dis, \p}^{\rm sym}\simeq  N_c \(\frac{c_\psi^2 y_\psi^2  }{2 \pi^3 f_\f^2}\) T^3. 
\eeq
The tau and charm Yukawa interactions give the dominant contribution to these processes.

{The most of the initial ALP energy is transferred to thermal plasma, and reheating is successful,  if the dissipation rate is larger than the Hubble parameter,}
\beq
\G_{\rm dis, \p}^{\rm sym} \gtrsim H_{\rm inf}
\eeq
by using $\frac{107.75 \pi^2}{30} T^4= V_0$ to estimate the dissipation rate. The relativistic degrees of freedom $1+106.75=107.75$ will be explained shortly.
{Let us define the ratio, $\xi \equiv H_{\rm inf}/\G_{\rm dis, \p}^{\rm sym}$.
In the ALP miracle scenario, some of the ALP condensate remains and explains DM. For this scenario, we need $\xi =\O(0.01-0.1)$~\cite{Daido:2017tbr}.}
Then we obtain 
\beq
\laq{reh}
f_\f/c_{5}^2=f_\f/c_3^2\simeq  5.8\times 10^7\GEV \frac{\xi}{0.1}\sqrt{\frac{3}{n}} \(\frac{\l}{10^{-12}}\)^{1/4}.
\eeq
{Since the photon anomaly coefficient, $c_\gamma$, satisfies, $c_\g\equiv c_{5} (1+ 5/3-1.92)$,} the photon coupling is
\beq
 g_{\f \g\g}\simeq 1.5\times 10^{-11}\GEV^{-1} c_3^{-1}\sqrt{\frac{n}{3}} \(\frac{\xi}{0.1}\)^{-1} \(\frac{\l}{10^{-12}}\)^{-1/4}.
\eeq
From \Eqs{ALPmass} and \eq{reh},
{we obtain a lower bound on the ALP mass $m_\phi$, which is relevant to determine the strong CP phase}
\beq
m_\f \gtrsim \sqrt{\d_{\rm CPV} m^2_\f} \sim 0.86\EV c_3^{-1/3} \(\frac{n}{3}\)^{1/6}\(\frac{g_{\f \g\g}}{1.5\times 10^{-11}\GEV^{-1}} \frac{\theta_{\rm QCD}}{10^{-10}}\)^{1/3} \(\frac{\l}{10^{-12}}\)^{1/6}.  \laq{lowboundmass}
\eeq
{In order to achieve successful reheating, i.e. the dominant fraction of the ALP condensate is thermalized, it is predicted that the ALP particle will undergo thermalization and become an additional degree of freedom due to its thermalization rate being almost identical to that given in \Eq{disH} \cite{Salvio:2013iaa}. This is why we have a hot DM component, which results in \Eqs{fphi} and \eq{strongCP}.
Again, we have checked that requiring the suppression of the hot DM density, $m_\phi<7.7\EV$, and \Eq{lowboundmass} predicts the small strong CP phase.}

So far we have focused on the perturbative effect of reheating.
In fact, there may also be a non-perturbative effect, namely the sphaleron process {, 
which could introduce friction to the ALP oscillation, potentially decreasing its amplitude, thereby converting the condensate energy into thermal plasma.}
Clarifying its precise contribution {to the reheating} is beyond the scope of this work, but we would like to comment on some specific features of our scenario.
Due to the negative curvature around the hilltop and large quartic coupling around the potential minimum, a tachyonic resonance and self-resonance occur soon after the inflation, before the end of the reheating process. As a result, a dominant fraction of the inflaton condensate is destroyed within a few oscillations of the ALP field. The resulting ALP condensate has typical energy of $p_{\rm typical}\sim \O(1-10)m_{\rm eff}\sim 10^{-3} \text{-}10^{-1}\L$~\cite{Lozanov:2017hjm}.\footnote{We use the term ``condensate" to refer to both the zero mode and excited modes due to the self-resonance.}
On the other hand, the sphaleron effect is characterized by the magnetic screening scale, $m_{\rm mag}= \O(0.1)g_i^2 T\sim 10^{-2}\L$ (see, e.g., Ref.\cite{Moore:2000ara}), where we have used $g_\star \pi^2/30 T^4=V_0$. Above this scale, the sphaleron rate quickly approaches zero\cite{Moore:2000ara}. Therefore, the interaction of excited modes with higher center-of-mass energy than $m_{\rm mag}$ may be irrelevant.
In our case, $m_{\rm mag}$ is comparable or slightly smaller than the typical center-of-mass energy between the ALP condensate and the thermal plasma, $\sqrt{T p_{\rm typical}}= \O(0.01-0.1)\L.$ Therefore, we expect a suppression of the QCD sphaleron rate from the conventional value of $\sim c_3^2 (3\a_3)^5 T^3/f_\f^2$. If the suppression is smaller than $\O(0.01)$, our conclusions do not change. 
The electroweak sphaleron effect is completely neglected in our analysis. However, we stress that a more detailed study of the sphaleron rate will be important.

\begin{figure}[!t]
\begin{center}  
\includegraphics[width=145mm]{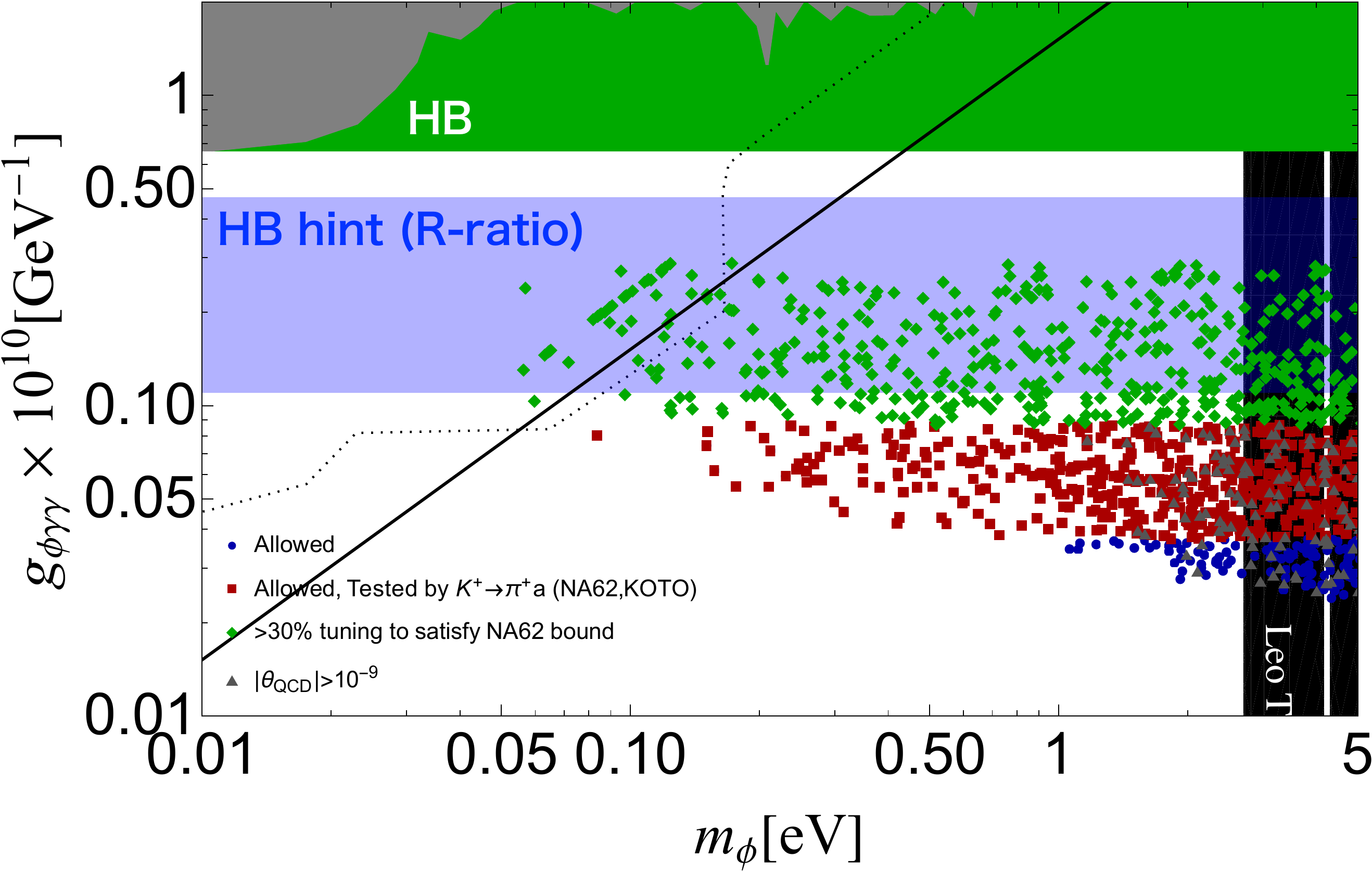}     
      \end{center}
\caption{The viable parameter region of the ALP being the inflaton and DM, featuring GUT (red square and blue circle data points), is shown in the $m_\f -g_{\f\g\g}$ plane. The black-shaded region indicates the bound from MUSE observations of the Leo T dwarf galaxy \cite{Regis:2020fhw}. {The red squares and green diamonds (with an amount of tuning less than 30\% to explain the current bound)} will also be tested in future measurements of $K\to\pi a$ in NA62 and KOTO experiments. The bound for the same process is around the HB's one. The gray triangle points denote the data with $\ab{\theta_{\rm QCD}}>10^{-9}$. Other constraints and hints are the same as in Figure \ref{fig:1}.
}\label{fig:ALP} 
\end{figure}

The scatter plot in Fig.\ref{fig:ALP} shows the parameter region for successful reheating via \Eq{lagreh}. We generated data points with varying values of $\log_{10}{\l}=[-13,-11]$, $\xi=[0.01,1]$, $\log_{10}{\z}=[-2,0]$, $\log_{10}|\theta_{\rm QCD}|=[-19, -5]$, $c_{5}=c_3=1,2,\cdots 10$, and $n=3,5,7,9,11$ at random. Using \Eqs{reh} and \eq{ALPmass}, we obtained $f_\f$ and $m_\f$ for each data point and plotted their position in the $m_\f-g_{a\g\g}$ plane. {We also checked the condition \eq{SNmu}, but found that it did not affect the viable parameter region.}  {\Eqs{Kpia} affected the parameter region. This bound can be alleviated slightly with a mild tuning by introducing the flavor-violating axion-quark coupling at the beginning to cancel the CKM bound. We show the green diamond points for the amount of tuning less than $30\%$ to have the allowed region, otherwise they are excluded.}
The red square points {will} be covered in future observations of $K^+\to \pi^+a$ via the CKM contribution. 
The blue circle data points may not be reached via $K^+ \to \pi^+ a$ experiments but can be fully tested in the future measurement of $\D N_{\rm eff}$ via CMB and BAO experiments, as well as via direct and indirect detections, including intensity mapping\cite{Baryakhtar:2018doz,Bessho:2022yyu,Shirasaki:2021yrp}, and laser-based collider experiments~\cite{Homma:2022ktv} via the ALP photon coupling. The gray diamond points denote the data with $|\theta_{\rm QCD}|>10^{-9}$.

Therefore, we have demonstrated that in order to explain both inflation and DM with an ALP featuring GUT, the parameter region should satisfy $g_{\f \gamma\gamma}=\O(10^{-11})\GEV^{-1}$, $0.01\EV\lesssim m_\f\lesssim 1\EV $, and $|\theta_{\rm QCD}|<10^{-9}.$ This conclusion is partly due to the fact that the predicted range from cosmology happens to be around the QCD axion parameter region, which depends on the size of the QCD scale, i.e., {the Hubble scale and the corresponding QCD-induced mass scale for the decay constant are similar.}
If the QCD scale were much larger than ours (but much smaller than $\L$), we could not find a parameter region for the ALP to serve as both the inflaton and DM (see Figs.\ref{fig:QCD}).\footnote{We could cancel $V_{\rm QCD}$ and $V_{\rm inf}$ contributions to be within \Eq{massrel}, but then the decay constant would be too large to achieve successful reheating.} If the QCD scale were much smaller than ours, then the strong CP phase would not be suppressed to the same extent.
The predicted viable parameter region is consistent with the bounds from stellar cooling, flavor physics, structure formation for the late-forming DM~\cite{Daido:2017wwb,Daido:2017tbr}, and the neutron EDM.
This ALP DM scenario is even hinted at by analyses of stellar cooling and extragalactic background light~\cite{Gong:2015hke, Kohri:2017oqn, Kalashev:2018bra, Korochkin:2019qpe, Caputo:2020msf}.

\section{Conclusions}
In this paper, we {have} investigated the possibility of realizing the hadrophobic axion in the context of GUT. In particular, we {have shown} that the required PQ charge assignment can be understood on the basis of isospin symmetry.
 If the hadrophobic axion is the QCD axion, 
we get the photon coupling as  the conventional GUT axion. 
  
  Furthermore we {have} imposed the condition for electrophobia to satisfy the severe astrophysical bound on the axion-electron coupling. Based on both conditions for the hadrophobia and electrophobia we have classified possible PQ charge assignments that are consistent with the SU(5) GUT.
The lower end of the axion window for the QCD axion 
can be slightly {relaxed} due to the hadrophobic and electrophobic nature.
This revives the region where the axion can explain some stellar cooling hints simultaneously. 
This interesting region can be fully tested in the future. 
There are also implications for the flavor physics, since it enhances the branching fraction of $K^+\to \pi^+ +\,$missing which can be observed in the KOTO and NA62 experiments.
Most of the viable parameter regions where the QCD axion is the dominant component of DM will be tested in the future haloscope and lumped element experiments.

Finally, we have studied a scenario, where an ALP plays both roles of the inflaton and DM, and shown that it is compatible with a viable model of the GUT EFT. Requiring the ALP as the inflaton and DM, the viable parameter region points to 
\beq
g_{\f \gamma\gamma}=\O(10^{-11})\GEV^{-1}, 0.01\EV\lesssim m_\f\lesssim 1\EV \AND |\bar\theta_{\rm CP} ~{\rm mod}~\pi|\lesssim 10^{-9}.
\eeq
Interestingly, in most of the above parameter regions, the strong CP problem is solved, and this scenario can be fully tested not only from the K-meson decay and star coolings, but also from the future measurement of the $\D N_{\rm eff}$, direct/indirect detections, {(the absence of) the primordial tensor mode}, and EDM.  

Once the axion DM is found, it could further become a probe of the origin of GUTs and flavors by measuring the axion couplings with fermions and photons.
\\
\\

{\it Note added:} While {this paper was being submitted}, Ref.~\cite{Badziak:2023fsc} appeared on arXiv. The authors discussed the (non-GUT) hadrophobic axion with the emphasis on the quantized charge assignment of \Eq{hadrophobic} as well as the interpretation by using the isospin conservation. {In fact, the same argument was already pointed out by one of the authors (WY)} on 9th February 2022
at ``The 2022 Chung-Ang University Beyond the Standard Model Workshop". The talk slides are available from:\url{
https://indico.cern.ch/event/1108846/contributions/4679286/}.

\section*{Acknowledgments}
We thank Kohsaku Tobioka for useful discussion on flavor experiments in a different context when he visited Tohoku University.  
This work is supported by JSPS Core-to-Core Program (grant number: JPJSCCA20200002) (F.T.),  JSPS KAKENHI Grant Numbers 20H01894 (F.T.), 20H05851 (F.T. and W.Y.),  21K20364 (W.Y.),  22K14029 (W.Y.), and 22H01215 (W.Y.). This article is based upon work from COST Action COSMIC WISPers CA21106,  supported by COST (European Cooperation in Science and Technology).

\appendix 

\section{Hadronic ambiguity and star coolings}
\lac{bound}

In this appendix, we take $c_u=2/3 c_3, c_d=1/3 c_3$ and use \Eq{precise} to estimate the star cooling bounds. 
As we will discuss, the hadronic ambiguity is important in the case that the isospin symmetry is an approximate good symmetry and $a$-nucleon coupling is zero-consistent. 
As long as the hadronic ambiguity dominates, our result will not change much if $c_u, c_d$ differ slightly from the sample value. Moreover, 
the bound we will derive is not very sensitive to the astrophysical bound we use because of the dominant hadronic ambiguity.

One bound on the nucleon coupling arises from the duration of the neutrino burst from SN1987A. The conservative analytical formula for the bound is given by~\cite{Zyla:2020zbs}:
\beq
\laq{SN}
g_{an}^2+0.29g_{ap}^2+0.27g_{an} g_{ap}\lesssim 3.25\times 10^{-18}.
\eeq
Using the central value $c_N=-0.02$ for the hadrophobic axion, we obtain a lower bound on the decay constant:
\beq 
 \sqrt{2}v_a/c_3\gtrsim 1.3 \times 10^7\GEV ~~(\text{SN1987A, mean}).
 \eeq
However, in our case, $c_N$ is zero-consistent and the value is dominated by hadronic uncertainty, which we have to take into account. To deal with the hadronic ambiguity, we introduce a probability distribution:
\beq
P[\bar{c}_N-\D c_N/2<c_N< \bar{c}_N+\D c_N/2  ]\simeq  f[c_n=0,c_p=0] \D c_n  \D c_p\sim 11^2 \D c_n \D c_p.
\eeq
Here we assume that $\bar{c}_n$ and $\bar{c}_p$ are close to zero and $(0<)\D c_N\ll \sigma_{c_N}.$ 
We further assume a normal distribution $f[c_n, c_p]$ with a mean value of $\mu_{c_N}=-0.02$ and a variance of $\sigma_{c_N}=0.03$ {as in \Eq{precise}}. For instance, we can find $P[|c_p| < 0.01, |c_n| < 0.01]\simeq 5\%$ by integrating $c_n$ and $c_p$ over the given range.
If we assume that the axion contribution is negligible when the l.h.s of \eq{SN} is smaller than half of the r.h.s, i.e., $1.63\times 10^{-18}$, then we can estimate the probability in the area of the ellipse with l.h.s $< 1.63\times 10^{-18}$ to find the decay constant in the $95\%$ region:
\begin{equation}
\sqrt{2}v_a/c_3 \gtrsim 6.1\times 10^6\,\mathrm{GeV} ~~\text{(SN1987A, hadronic ambiguity)}.
\end{equation}

Next we focus on the cooling of NS in the supernova remnant of Cassiopeia A (Cas A). 
It was known that the cooling rate was measured to be anomalously fast for certain models of NS.
Such a rapid cooling can also be explained by neutron superfluidity and proton superconductivity~\cite{Page:2010aw, 10.1111/j.1745-3933.2011.01015.x} or a phase transition of the neutron condensate~\cite{Leinson:2014cja}.
Assuming that the minimal model explains the cooling of the Cas A, 
we obtain a constraint on the axion coupling~\cite{Hamaguchi:2018oqw} (see also \cite{Leinson:2021ety})
$
\laq{CASA}
g_{ap}^2+1.6g_{an}^2\lesssim  10^{-18}.
$
By using the central value we obtain 
\beq
\sqrt{2}v_a/c_3\gtrsim 3\times 10^{7}\GEV~~~\text{(Cas A, mean)}.
\eeq
Again by requiring the l.h.s $< 0.5\times 10^{-18}$, and by including the hadronic ambiguity we obtain with $95\%$ probability  
$
\sqrt{2}v_a/c_3 \gtrsim 1.2\times 10^7\GEV~~~ \text{(Cas A, hadronic ambiguity)}. 
$
Several other bounds on individual $g_{ap}$ or $g_{an}$ can also be found.  
A conservative bound~\cite{Sedrakian:2015krq} is $g_{ap}^2 \lesssim (1\text{-}6)\times 10^{-17}$ by neglecting the transient behavior of Cas A. 
For the mean value this leads to $\sqrt{2}v_a/c_3\gtrsim 6\times 10^6\GEV.$ Including the ambiguity it is much smaller. 

An interesting hint for cooling was reported in~\cite{Leinson:2014ioa} (see also~\cite{Hong:2020bxo}), with $g_{an}^2 = (1.4 \pm 0.5) \times 10^{-19}$. To estimate the probability distribution of $\sqrt{2}v_a$, we assume that the hint is represented by a normal distribution with a peak at $1.4 (0.5) \times 10^{-19}$ and take into account the hadronic ambiguity. The preferred range for $\sqrt{2}v_a/c_3$ is found to be between $4.0 \times 10^{7}\GEV$ and $1.9\times 10^8\GEV$, corresponding to the limits of half the peak of the distribution. This range is consistent with the cooling limit of SN1987A, whether or not the hadronic ambiguity is included. However, the hint may be affected by a longer period of Cas A temperature measurements and further understanding of the cooling mechanism~\cite{Hamaguchi:2018oqw, Leinson:2021ety}. Therefore, we do not apply this hint in the main part, but we expect that this region may be probed by Cas A cooling in the future.

Let us consider the cooling of HB stars.
The $R$ parameter sets a bound of~\cite{Raffelt:1985nk,Raffelt:1987yu,Raffelt:1996wa, Ayala:2014pea, Straniero:2015nvc, Giannotti:2015kwo, Carenza:2020zil} $g_{a\g\g}\lesssim 6.6 \times 10^{-11}\GEV^{-1}$ (95\% CL), by assuming no significant axion contribution to the red giant stars (see also a stronger bound in \cite{Dolan:2022kul}).
If there is no additional axion-photon coupling, $g_{a\g\g}\simeq g_{a \g\g}^{a-\pi}$. Using the central value, we obtain
\beq
\sqrt{2} v_a/c_3 \gtrsim 1.4\times 10^6\GEV ~~(\text{HB: no extra photon coupling}).
\eeq
An extra photon coupling may arise,
\beq
\d{\cal L}\supset -c_\g \frac{e^2 a }{32\pi^2 \sqrt{2} v_a} F\tl F
\eeq
in addition to \Eq{lag}.
In this case, we have $g_{a\g\g}\simeq c_\g \frac{e^2}{8\pi^2 \sqrt{2} v_a}$, which dominates over the suppressed pion contribution for $c_\g=\O(1)$. This gives
\beq
\sqrt{2} v_a \gtrsim ~~c^{-1}_\g 1.7\times 10^7\GEV  (\text{HB: with extra photon coupling}).
\eeq
If a tree-level axion-gauge coupling is given, the 2-loop RG running induces the derivative coupling to fermions~\cite{Chakraborty:2021wda}. In the region of $\sqrt{2} v_a\gtrsim 10^7\GEV$, this contribution can be neglected in the context of cooling for stars (including the red giant branch stars).
On the contrary, there is a hint from the $R$ parameter~\cite{Ayala:2014pea, Straniero:2015nvc, Giannotti:2015kwo}
\beq \laq{HB}g_{a\g\g} = ~~\(0.29 \pm 0.18\) \times10^{-10}\GEV^{-1} (\text{HB hint}).\eeq  
We refer to this as the HB hint.

The muon coupling is constrained to be 
\beq \laq{SNmu}g_{a\mu\mu}\equiv \frac{c_\mu m_\mu}{\sqrt{2} v_a} \lesssim 2\times10^{-9}\eeq from SN1987A~\cite{Bollig:2020xdr,Croon:2020lrf}, which do not suffer much from the hadronic ambiguity.

\bibliography{draft5.bib}
\end{document}